\documentclass{aa}  

\usepackage{graphicx}
\usepackage{txfonts}
\usepackage{subfig}
\usepackage{hyperref}
\usepackage{color}

\begin{document}

   \title{The GAPS programme at TNG\thanks{Based on: observations made with the Italian \textit{Telescopio Nazionale Galileo} (TNG), operated on the island of La Palma by the INAF - \textit{Fundaci\'on Galileo Galilei} at the \textit{Roque de Los Muchachos} Observatory of the \textit{Instituto de Astrof\'isica de Canarias} (IAC).}}

   \subtitle{LXX. HD\,128717\,B/Gaia-6\,B: A long-period eccentric low-mass brown dwarf from astrometry and radial velocities}

   \titlerunning{Gaia-6\,B}

   \author{M. Pinamonti\inst{\ref{inst1}}
          \and
          A. Sozzetti\inst{\ref{inst1}}
          \and
          D. Barbato\inst{\ref{inst2}}
          \and
          S. Desidera\inst{\ref{inst2}}
          \and
          K. Biazzo\inst{\ref{inst3}}
          \and
          A. S. Bonomo\inst{\ref{inst1}}
          \and
          A. F. Lanza\inst{\ref{inst6}}
          \and
          L. Naponiello\inst{\ref{inst1}}
          \and
          L. Affer\inst{\ref{inst11}}
          \and
          R. M. Anche\inst{\ref{steward}}
          \and
          G. Andreuzzi\inst{\ref{inst12},\ref{inst3}}
          \and
          M. Basilicata\inst{\ref{inst1}}
          \and
          M. Brinjikji\inst{\ref{sese}}
          \and
          M. Brogi\inst{\ref{inst1},\ref{inst9}}
          \and
          L. Cabona\inst{\ref{inst16}}
          \and
          E. Carolo \inst{\ref{inst2}}
          \and
          S. Colombo\inst{\ref{inst11}}
          \and
          M. Damasso\inst{\ref{inst1}}
          \and
          M. D’Arpa\inst{\ref{inst11}}
          \and
          S. Di Filippo \inst{\ref{inst2}}
          \and
          A. Harutyunyan\inst{\ref{inst12}}
          \and
          J. Hom\inst{\ref{steward}}
          \and
          L. Mancini\inst{\ref{inst4},\ref{inst1},\ref{inst5}}
          \and
          G. Mantovan\inst{\ref{inst8},\ref{inst2}}
          \and
          D. Nardiello\inst{\ref{inst7},\ref{inst2},\ref{inst8}}
          \and
          K. K. R. Santhakumari \inst{\ref{inst2}}
          \and 
          T. Zingales\inst{\ref{inst7},\ref{inst2}}
          }

   \institute{INAF - Osservatorio Astrofisico di Torino, Via Osservatorio 20, I-10025 Pino Torinese, Italy\\
              \email{matteo.pinamonti@inaf.it}\label{inst1}
          \and
              INAF - Osservatorio Astronomico di Padova, vicolo dell'Osservatorio 5, I-35122 Padova, Italy\label{inst2}
         \and
             INAF - Osservatorio Astronomico di Roma, Via Frascati 33, I-00078 Monte Porzio Catone, Italy\label{inst3}
         \and
             INAF - Osservatorio Astrofisico di Catania, Via S. Sofia 78, I-95123 Catania, Italy\label{inst6}
         \and
             INAF - Osservatorio Astronomico di Palermo, piazza del Parlamento 1, I-90134 Palermo, Italy\label{inst11}
        \and
            Department of Astronomy and Steward Observatory, The University of Arizona, 933 North Cherry Ave, Tucson, AZ85721, USA \label{steward}   
         \and
             Fundaci\'on Galileo Galilei - INAF, Ramble Jos\'e Ana Fernandez P\'erez 7, E-38712 Bre\~na Baja, TF, Spain\label{inst12}
         \and
            School of Earth and Space Exploration, Arizona State University, 781 E Terrace Mall, Tempe, AZ 85287, USA\label{sese}
         \and
             Department of Physics, University of Turin, Via Pietro Giuria 1, 10125 Torino, Italy\label{inst9}
         \and
             INAF - Osservatorio Astronomico di Brera, Via E. Bianchi 46, I-23807 Merate, Italy\label{inst16}
         \and
             Dipartimento di Fisica, Universit\`a di Roma ``Tor Vergata'', Via della Ricerca Scientifica 1, I-00133 Roma, Italy\label{inst4}
         \and
             Max Planck Institute for Astronomy, Kõnigstuhl 17, DE-69117 Heidelberg, Germany\label{inst5}
         \and
             Dipartimento di Fisica e Astronomia, Universit\`a degli Studi di Padova, Vicolo dell’Osservatorio 3, 35122 Padova, Italy\label{inst7}
         \and
             Centro di Ateneo di Studi e Attività Spaziali “G. Colombo” – Università degli Studi di Padova, Via Venezia 15, 35131 Padova, Italy\label{inst8}
         \and
             INAF - Osservatorio Astronomico di Brera, Via E. Bianchi 46, I-23807 Merate, Italy\label{inst10}
         }

   \date{---}

 
  \abstract
   {The transition regime between giant planets (GPs) and brown dwarfs (BDs) is still an open subject of study in exoplanetary science. A complete understanding of the  population of long-period GPs and BDs would be pivotal for improving our knowledge of this topic, but the number of such objects with precisely measured orbital and physical parameters remains small. Moreover, their dynamical influence on smaller companions in inner orbits is still unclear.}
   {Within the framework of the Global Architecture of Planetary Systems collaboration (GAPS), we aim to confirm and characterise sub-stellar companion candidates from Gaia DR3, and to study the potential presence of additional lower mass planets in their systems.}
   {We present the results of an intensive high-precision radial velocity (RV) monitoring of HD\,128717, which hosts the astrometric candidate Gaia-ASOI-009. We used the HARPS-N spectrograph at Telescopio Nazionale Galileo (TNG) to collect a high-cadence RV time series of the target. We used Markov chain Monte Carlo (MCMC) analyses to refine the Gaia DR3 orbital solution of the companion and, finally, performed a combined  model of RV and proper motion anomaly (PMa) to derive the complete 3D orbit of the companion. We also ran a suite of numerical simulations to confirm our results.}
   {We confirmed the sub-stellar nature of Gaia-ASOI-009 (i.e. Gaia-6\,B). From the combined RV+PMa fit, we confirm that it is a high-eccentricity low-mass BD with $P_\text{B} = 9.37^{+0.06}_{-0.05}$ yr, $M_\text{B} = 19.8 \pm 0.5$ $M_\text{J}$, and $e_\text{B}=0.85$, $i_\text{B} = 130^{\circ}$.
   The derived orbital solution differs significantly from the one published in Gaia DR3. Through a series of dedicated simulations, we demonstrated that this discrepancy arises from a degeneracy in the Gaia DR3 astrometric solution. Specifically, the combination of Gaia-6\,B long orbital period and high eccentricity, both poorly constrained by the limited time span of DR3, led to an incorrect solution characterised by a shorter period and lower eccentricity. Finally, we find no evidence of other companions in the system (in the inner or the outer regions). Thus, the origin of the high-eccentricity of Gaia-6\,B remains unclear.}
   {}

   \keywords{planetary systems - techniques: spectroscopic - Astrometry - techniques: radial velocities - stars: individual: HD\,128717 – methods: data analysis
               }

   \maketitle
%

\section{Introduction}

    In the study of large sub-stellar companions, one of the main questions is whether the object can be classified as an extrasolar planet or as a brown dwarf (BD). In the transition regime, roughly between $10$ and $20$ $M_\text{J}$, providing a definitive answer to this question can be very challenging. Traditionally, the threshold between the two types of objects has been set at 13 $M_\text{J}$, which is considered to be the limiting mass for the fusion of deuterium \citep{bossetal2003}. However, more recent studies have highlighted that the lower-mass limit for deuterium burning might vary between 11 and 16 $M_\text{J}$, depending on the object composition and other factors \citep[e.g.][]{mollieremordasini2012,schneider2018}. Other studies have suggested using the formation mechanisms to distinguish the two populations of companions \citep[e.g.][]{chabrieretal2014}, defining  the companions formed via gravitational instability as BDs \citep[e.g.][]{bosskanodia2023}, while exoplanets are considered to be those formed via  core-nucleated or pebble-assisted accretion within a disc \citep[e.g.][]{mordasinietal2012b}. Unfortunately, these definitions are difficult to apply with respect to observed companions because the formation history is difficult to infer from the observed properties, since different formation models can produce similar outcomes. Moreover, the number of currently known sub-stellar companions on the transition regime between BD and GP with precisely measured orbital parameters is still very limited. Most of the currently known long-period objects come from radial velocity (RV) surveys, which means that the true mass is unknown, making it extremely difficult to unveil their true nature.

The advent of ESA’s Gaia mission \citep{gaiaetal2016} provides a promising new avenue for overcoming these limitations. Gaia's astrometric capabilities enable precise measurements of stellar positions and motions, making it a powerful tool for detecting the gravitational influence of sub-stellar companions, especially in intermediate to long-period orbits. In particular, astrometric data from Gaia complement RV measurements by offering an independent method for determining the true mass and orbital parameters of companions. The extended Gaia mission now holds the potential to detect GPs and low-mass BDs out to distances of 3-5 AU, for stars of 0.3-1 $M_\odot$, a critical regime for understanding the transition between these two classes of objects. The first astrometric results from Gaia \citep{gaiavallenarietal2023}, including detections of sub-stellar companions, are already providing valuable insights into the demographics of these systems, with a few thousand orbits of sub-stellar companion candidates having been published as part of Gaia Data Release (DR) 3 \citep{gaiaarenouetal2023}. This unprecedented dataset, combined with complementary RV data, sets the stage for a new phase in exoplanet research, offering a more complete view of system formation and evolution.

The first two confirmed Gaia exoplanets were transiting hot-Jupiters detected in the photometry, Gaia-1b and Gaia-2b \citep{panahietal2022}.
Subsequently, \citet{sozzettietal2023} announced the confirmation of the first astrometric Gaia candidate, Gaia-3b, a super-Jupiter on a 300 d period orbit.
However, later internal investigation of the Gaia pipeline revealed that the time-series for this target suffered from a software bug and that the orbital solution published in Gaia DR3 was a false positive\footnote{\url{https://www.cosmos.esa.int/web/gaia/dr3-known-issues\#FalsePositive}}.
Since the RV planetary signal has been confirmed, Gaia-3b still represents a Gaia-enabled discovery, as the target would likely not have been monitored without its initial identification as an astrometric candidate \footnote{This, however, should be regarded as a statistical exception rather than an expected outcome.}.
Finally, \citet{stefanssonetal2025} announced the confirmation of two more astrometric Gaia planet candidates, Gaia-4b and Gaia-5b, identified as a massive planet (11.8 $M_\text{J}$) at $P = 571.3$ d and a low-mass BD (20.9 $M_\text{J}$) at $P = 358.58$ d.
These companions were confirmed by combining RV measurements from the Habitable-zone Planet Finder (HPF), NEID, and FIES spectrographs. 

In this context, we present results from an intensive high-precision RV monitoring campaign of another Gaia astrometric planet candidate, Gaia-ASOI-009 (aka HD\,128717). The object is part of the publicly available Gaia candidate exoplanet list\footnote{\url{https://www.cosmos.esa.int/web/gaia/exoplanets}}. The Gaia DR3 archive reports a value of the renormalised unit
weight error (RUWE)\footnote{ A standard diagnostic of the departure from a good single-star fit to Gaia astrometry.} statistic of 1.32, formally below the threshold of 1.4 adopted for systematic attempts to perform orbital fits to Gaia DR3 astrometric time-series. The Gaia DR3 solution type, `OrbitalTargetedSearch', corresponds to one of the orbital solutions obtained as part of an analysis performed on a supplementary external input list, as described in detail in \citet{holletal2023} and \citet{gaiaarenouetal2023}. We observed HD\,128717 within the context of the Global Architecture of Planetary Systems collaboration \citep[GAPS,][]{covinoetal2013,desideraetal2013} with the HARPS-N spectrograph \citep{cosentinoetal2012} at Telescopio Nazionale Galileo (TNG). The target was included in the target list of the original GAPS programme, with a tailored observing strategy designed upon release of the Gaia DR3 orbital solution. HD\,128717 was then included in the ongoing large programme: The Great HARPS-N hunt for super-Earths and Neptunes interior to outer giant planets detected by Gaia. This programme is focussed on confirming astrometric long-period exoplanets and BDs and, through subsequent high-cadence follow-up, as well as on searching for inner low-mass planets in the observed systems. A full description of the programme will be provided in an upcoming paper (Barbato et al., in prep.).
Here, we present a detailed analysis of the RV dataset collected within the programme, as well as a combined analysis with the available Gaia DR3 and Hipparcos astrometry to confirm the orbital solution, along with a set of simulations to confirm the planetary nature of the candidate.

In Sects.  \ref{sec:spec_data} and \ref{sec:star_par}, we present the new stellar parameters derived for the target from the HARPS-N spectra and the RV data collected for this study, respectively. In Sect. \ref{sec:kep_fit}, we present the Keplerian fits of the RV time series and the combined analysis of RVs and astrometric data. In Sect. \ref{sec:gaia_sim}, we present our simulations of Gaia observations. In Sect. \ref{sec:pma_fit}, we combine the PMa astrometric information with our RV data to perform a joint modelling of the orbit. In Sect. \ref{sec:discussion}, we discuss our results and the follow-up imaging observations we performed. Finally, in Sect. \ref{sec:conclusion} we summarise our findings.

\section{Spectroscopic HARPS-N data}
\label{sec:spec_data}

As part of the GAPS RV programme, HD\,128717 was observed with HARPS-N from BJD $= 2459735.6$ (5 June 2022) to BJD $= 2460686.7$ (11 January 2025). In total, 106 spectra were collected over a time span of 951 days.
These spectra were reduced with version 3.7.1 of the HARPS-N DRS (data reduction software) pipeline, maintained by the Italian Centre for Astronomical Archive (IA2)\footnote{\url{https://ia2.inaf.it}}. 
All observations were collected with an integration time of 600 s, to average out potential short-term periodic oscillations of the star \citep{dumusqueetal2011} and to obtain a sufficient per-pixel signal-to-noise ratio (average S /N = 40 at 550 nm).

\begin{figure}
   \centering \includegraphics[width=0.45\textwidth]{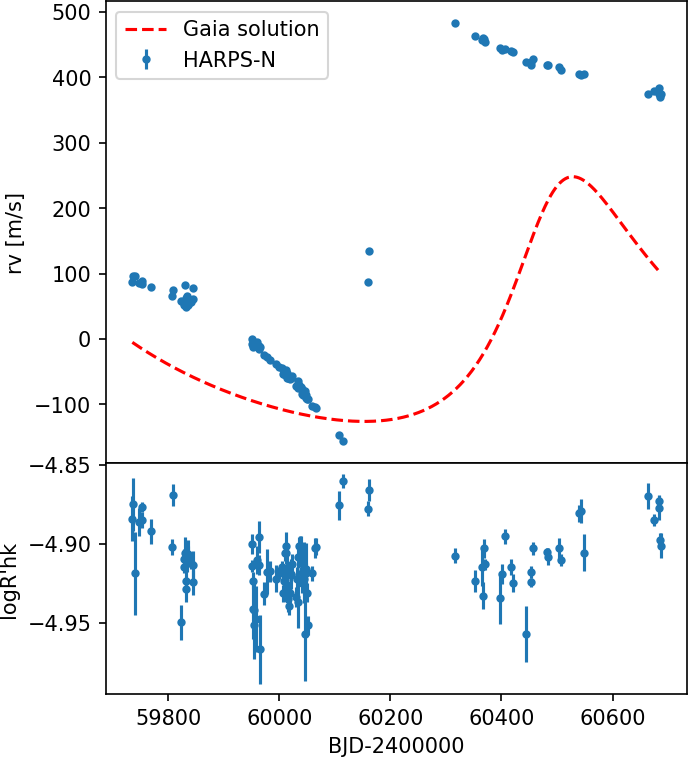}
      \caption{\textit{Upper panel:} HARPS-N RV time series of HD\,128717 (blue points), compared with the predicted RV signals from the Gaia orbital solution (red dashed line). \textit{Lower panel:} HARPS-N $\log R'_\text{HK}$ time series.}
         \label{fig:rv_TERRA_Gaia}
\end{figure}

The RV measurements were derived using the Template Enhanced Radial velocity Reanalysis Application \citep[TERRA,][]{anglada-escudebutler2012}. The resulting RV time series has a median error of $\sigma_\text{RV} = 1.6$ m s$^{-1}$ and a rms of $196.7$ m s$^{-1}$. The RV dataset is shown in Fig. \ref{fig:rv_TERRA_Gaia}. The complete time series of all spectroscopic data used in this work is available at the CDS. This figure also displays the expected RV signal corresponding to the best-fit
orbital solution for HD\,128717\,b from Gaia DR3 \citep{gaiaarenouetal2023}.
The DR3 orbital parameters are listed in Table \ref{tab:gaia_sol}. The table
lists (along with the parameters available in the {\fontfamily{lmtt}\selectfont
gaiadr3.nss\_two\_\text{b}ody\_orbits} and {\fontfamily{lmtt}\selectfont
gaiadr3.binary\_masses} tables at the Gaia archive)\footnote{\url{https://gea.esac.esa.int/archive/}}
the derived value of the Campbell elements, along with the values of the
minimum mass, $m_\text{b} \sin{i}$, and RV semi-amplitude, $K_\text{b}$.
 
As we can see, there is a clear inconsistency between the predicted planetary
signal and the HARPS-N observations, both in amplitude and phase. It is worth
noticing that the orbital period listed in Table \ref{tab:gaia_sol} corresponds
to the 34 months time span of Gaia DR3. This might suggest an underestimated
period due to insufficient phase coverage and we discuss this discrepancy
in detail in the following sections.

\begin{table}
\caption{Gaia DR3 orbital solution for ASOI-009 from \cite{gaiaarenouetal2023}.}
\small
\label{tab:gaia_sol}
\centering
\begin{tabular}{lc}
\hline\hline
\noalign{\smallskip}
Parameter & HD\,128717\\
\hline
\noalign{\smallskip}
$M_\star$ $[$M$_\odot]$ &  $1.128^{+0.058}_{-0.160}$ \\
\noalign{\smallskip}
$M_\text{b}$ $[M_\text{J}]$ &  10.0  \\
\noalign{\smallskip}
$P_\text{b}$ $[\text{d}]$ &  $1090 \pm 310$  \\
\noalign{\smallskip}
$e_\text{b}$ &  $0.39 \pm 0.19$  \\
\noalign{\smallskip}
$t_\text{P} - 2016.0$ $[\text{d}]$ &  $-180 \pm 290$  \\
\noalign{\smallskip}
A $[$mas$]$ & $0.16 \pm 0.13$ \\
\noalign{\smallskip}
B $[$mas$]$ & $0.25 \pm 0.16$ \\
\noalign{\smallskip}
F $[$mas$]$ & $-0.25 \pm 0.18$ \\
\noalign{\smallskip}
G $[$mas$]$ & $0.18 \pm 0.18$ \\
\noalign{\smallskip}
\multicolumn{2}{c}{\textit{Campbell elements}}  \\
\noalign{\smallskip}
$a_0$ $[$mas$]$ & $0.47 \pm 0.14$ \\
\noalign{\smallskip}
$\omega$ $[$deg$]$ & $324 \pm 58$ \\
\noalign{\smallskip}
$\Omega$ $[$deg$]$ & $91 \pm 50$ \\
\noalign{\smallskip}
$i$ $[$deg$]$ & $67 \pm 19$ \\
\noalign{\smallskip}
\multicolumn{2}{c}{\textit{RV parameters}}  \\
\noalign{\smallskip}
$m_\text{b} \sin{i}$ $[M_\text{J}]$ & $8.7 \pm 1.6$ \\
\noalign{\smallskip}
$K_\text{b}$ $[$m s$^{-1}]$ & $185 \pm 53$ \\
\noalign{\smallskip}
\hline
\end{tabular}
\tablefoot{$M_\star$ and $M_\text{b}$ from {\fontfamily{lmtt}\selectfont gaiadr3.binary\_masses}; $P_\text{b}$, $e_\text{b}$, $t_\text{P}$, A, B, F, and G from {\fontfamily{lmtt}\selectfont gaiadr3.nss\_two\_\text{b}ody\_orbits}; Campbell elements and RV parameters derived here.}
\end{table}

\section{Stellar properties}
\label{sec:star_par}

We derived the effective temperature ($T_{\rm eff}$), the surface gravity ($\log g$), the microturbulence velocity ($\xi$), and the iron abundance ([Fe/H]) using the equivalent width method to the co-added spectrum of the target (see \citealt{Biazzoetal2022} for details). We considered the MOOG code (\citealt{Sneden1973}; version 2019), the iron line list by \cite{Biazzoetal2022}, and the ATLAS9+ODFNEW (\citealt{CastelliKurucz2003}) grids of model atmospheres. Briefly, $T_{\rm eff}$ was measured by imposing the excitation equilibrium of \ion{Fe}{i} lines, $\log g$ through the ionisation equilibrium between \ion{Fe}{i} and \ion{Fe}{ii} lines, and $\xi$\,was obtained by removing the trend between the \ion{Fe}{i} abundances and the reduced equivalent width $REW=\log(EW/\lambda)$.
From this, we found $T_{\rm eff} = 6320 \pm 40$ K, $\log g = 4.37 \pm 0.13$, and $[Fe/H] = 0.16 \pm 0.08$.

Moreover, adopting the macroturbulence velocity of 4.7\,km\,s$^{-1}$ by \cite{Doyleetal2014}, we also derived the projected rotational velocity $v\sin i_{\star}=6.1\pm0.5$\,km\,s$^{-1}$ via the spectral synthesis technique applied to three different wavelength regions of the co-added spectrum (see \citealt{Biazzoetal2022} for the procedure). 

We also detected the presence of the lithium line at 6707.8\,\AA, deriving an equivalent width of $67.0\pm1.0$ m\AA. Adopting our spectroscopic parameters measured above, we obtained a lithium abundance of $\log(N_{\rm Li})_{\rm NLTE}=2.78\pm0.02$\,dex, after correcting for non-local thermodynamical equilibrium (NLTE) effects (\citealt{Lindetal2009}). This high lithium abundance might suggest a young stellar age, discussed further in Appendix \ref{app:star_age}.

\begin{figure}
\centering
\includegraphics[width=1\linewidth]{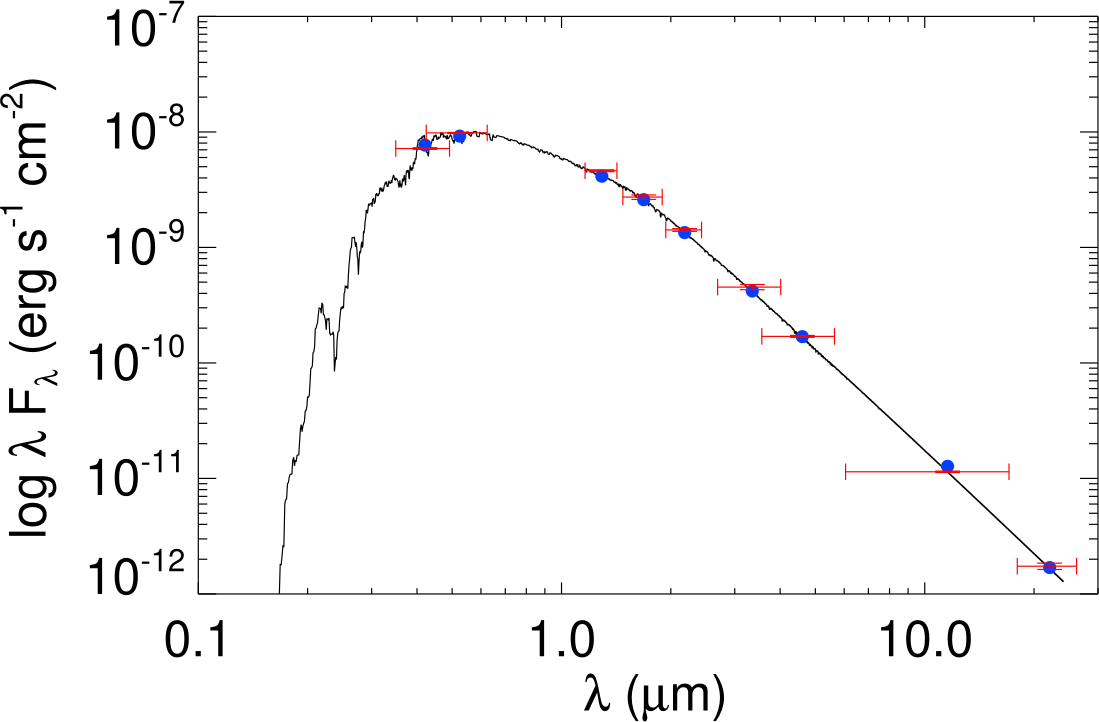}
\caption{Stellar spectral energy distribution (SED). The broad-band measurements from the Tycho, 2MASS, and WISE magnitudes are shown in red and the corresponding theoretical values with blue circles. The unaveraged best-fit model is displayed with a black solid line.} 
\label{fig:stellarSED}
\end{figure}

We determined the stellar mass, radius, and age by simultaneously modelling the stellar spectral energy distribution (SED) and the MESA Isochrones and Stellar Tracks \citep{paxtonetal2015} through the {\tt EXOFASTv2} differential evolution Markov chain Monte Carlo tool (MCMC; see \citealt{eastman2017, eastmanetal2019} and \citealt{naponielloetal2025} for more details). To sample the SED, we used the available Tycho-2 $B_{\rm T}$ and $V_{\rm T}$, 2MASS $J$, $H$, and $K_{\rm s}$, and WISE $W1$, $W2$, $W3$, and $W4$ magnitudes (see Table~\ref{tab:star_par} and Fig.~\ref{fig:stellarSED}). 
We imposed Gaussian priors on the $T_{\rm eff}$ and [Fe/H] from the spectral analysis above, as well as on the Gaia DR3 parallax. 
From the medians and $16^\text{th}-84^\text{th}$  percentiles of the posterior distributions we found  $M_\star=1.212^{+0.058}_{-0.068}~\rm M_\odot$, $R_\star=1.248 \pm 0.024~\rm R_\odot$, and age $t=1.9^{+1.9}_{-1.3}$~Gyr; the corresponding $\log g$ is fully consistent with the spectroscopic $\log g$ determined from the spectral analysis. All the parameters of the host star are given in Table~\ref{tab:star_par}. 

\begin{table}
\caption{Stellar parameters of HD\,128717 and orbital solution for HD\,128717 from Gaia DR3.}
\small
\label{tab:star_par}
\centering
\begin{tabular}{lcc}
\hline\hline
Parameter & HD\,128717 & Reference\\
\hline
\noalign{\smallskip}
\multicolumn{3}{c}{\textit{Astrometric parameters}}  \\
\noalign{\smallskip}
$\alpha$ (J2000) &  14$^h$:36$^m$:21.3$^s$ & (a) \\
\noalign{\smallskip}
$\delta$ (J2000) & +57$^\circ$:33$'$:38.4$''$ & (a)  \\
\noalign{\smallskip}
$\mu_\alpha$ $[\text{mas yr}^{-1}]$ & $-82.125\pm 0.022$ & (a) \\
\noalign{\smallskip}
$\mu_\delta$ $[\text{mas yr}^{-1}]$ & $67.259 \pm 0.025$ & (a) \\
\noalign{\smallskip}
$\pi$ $[\text{mas}]$ & $13.560 \pm 0.022$ & (a) \\
\noalign{\smallskip}
RUWE &  1.32 & (a)  \\
\noalign{\smallskip}
d &  $73.74 \pm 0.12$~pc & This work  \\
\noalign{\smallskip}
\hline
\noalign{\smallskip}
\multicolumn{3}{c}{\textit{Photometric parameters}}  \\
\noalign{\smallskip}
$B-V$ $[\text{mag}]$ & 0.627 & (b) \\
\noalign{\smallskip}
$B_{\rm T}$ $[\text{mag}]$ & $8.981 \pm 0.018$ & (b)  \\
\noalign{\smallskip}
$V_{\rm T}$ $[\text{mag}]$ & $8.354 \pm 0.012$ & (b)  \\
\noalign{\smallskip}
$G$ $[\text{mag}]$ & $ 8.2003 \pm 0.0028$ & (a) \\
\noalign{\smallskip}
$J$ $[\text{mag}]$ & $7.309 \pm 0.021$ & (c) \\
\noalign{\smallskip}
$H$ $[\text{mag}]$ & $7.078 \pm 0.049$ & (c) \\
\noalign{\smallskip}
$K$ $[\text{mag}]$ & $7.039 \pm 0.029$ & (c) \\
\noalign{\smallskip}
$W1$ $[\text{mag}]$ & $6.954 \pm 0.056$ & (d) \\
\noalign{\smallskip}
$W2$ $[\text{mag}]$ & $7.041 \pm 0.020$ & (d) \\
\noalign{\smallskip}
$W3$ $[\text{mag}]$ & $7.049 \pm 0.016$ & (d) \\
\noalign{\smallskip}
$W4$ $[\text{mag}]$ & $7.028 \pm 0.068$ & (d) \\
\noalign{\smallskip}
$A_{\rm V}$ $[\text{mag}]$ & $ < 0.038$ & This work \\
\noalign{\smallskip}
\hline
\noalign{\smallskip}
\multicolumn{3}{c}{\textit{Stellar parameters}}  \\
\noalign{\smallskip}
\noalign{\smallskip}
$T_{\text{eff}}$ (spec.) $[$K$]$ & $6210 \pm 40$ & This work \\
$\log g$ (spec.) $[$cgs$]$ & $4.37 \pm 0.13$ & This work \\
\noalign{\smallskip}
$V_\text{micro}$ $[$km s$^{-1}]$ & $1.22 \pm 0.05$ & This work \\
\noalign{\smallskip}
$[$Fe$/$H$]$ $[$dex$]$ & $0.16 \pm 0.08$ & This work \\
\noalign{\smallskip}
\noalign{\smallskip}
$v \sin{i_\star}$ $[$km s$^{-1}]$ & $6.1 \pm 0.5$ & This work \\
\noalign{\smallskip}
$EW_{Li}$  $[$m\r{A}$]$ & $67.0 \pm 1.0$  & This work \\
\noalign{\smallskip}
$\log (N_{\rm Li})_{NLTE}$ $[$dex$]$ & $2.78 \pm 0.02$  & This work \\
\noalign{\smallskip}
$M_\star$ $[M_\odot]$ & $1.212^{+0.058}_{-0.068}$ & This work   \\
\noalign{\smallskip}
$R_\star$ $[R_\odot]$ & $1.248 \pm 0.024$ & This work  \\
\noalign{\smallskip}
$\rho_\star$ $[$g cm$^{-3}]$ & $0.879^{+0.065}_{-0.068}$ & This work \\
\noalign{\smallskip}
$\log g$ (track) $[$cgs$]$ & $4.329^{+0.025}_{-0.030}$ & This work \\
\noalign{\smallskip}
$\log L_*/L_\odot$ & $0.322^{+0.014}_{-0.015}$ & This work \\
\noalign{\smallskip}
$\log R'_\text{HK}$ & $- 4.948 \pm 0.020$ & This work  \\
\noalign{\smallskip}
Age $[$Gyr$]$ & $1.4 \pm 0.3$ & Sect. \ref{app:star_age}  \\
\noalign{\smallskip}
\hline                                   
\end{tabular}
\tablefoot{\tablefoottext{a}{\citet{gaiavallenarietal2023}}; \tablefoottext{b}{\citet{hogetal2000}}; \tablefoottext{c}{\citet{cutrietal2003}}; 
\tablefoottext{d}{\citet{cutrietal2013}}}
\end{table}

\section{Spectroscopic data analyses}
\label{sec:kep_fit}

In this section, we discuss the Keplerian fits of the RV time series of HD\,128717. We also present the ancillary analyses of the spectroscopic activity indices, performed to confirm the Keplerian RV signal.
We first present a blind analysis of the spectroscopic data, as would be performed to discover and characterise the RV signal from an unknown planetary candidate. We then discuss the combination of the astrometric Gaia DR3 candidate solution and the spectroscopic information for a complete characterisation of the orbital parameters.

The identification of periodic signals in the RV (or activity) time series was performed using a generalised Lomb-Scargle periodogram \citep[GLS,][]{zechkur2009}. The Keplerian signal was then fully characterised using a Markov chain Monte Carlo (MCMC) approach, specifically employing the emcee Affine Invariant MCMC Ensemble sampler developed by \citet{foreman13}. We included offset and jitter terms to take into account instrumental effects, and also tested a multi-Keplerian model and a linear RV trend to evaluate the presence of additional companions in the system. The model comparison was performed via the Bayesian information criterion \citep[BIC,][]{schwarz1978}. The adopted analysis approach is the same as in \citet{pinamontietal2023}.

\subsection{Periodogram analyses}

\begin{figure}
   \centering
 \includegraphics[width=.45\textwidth]{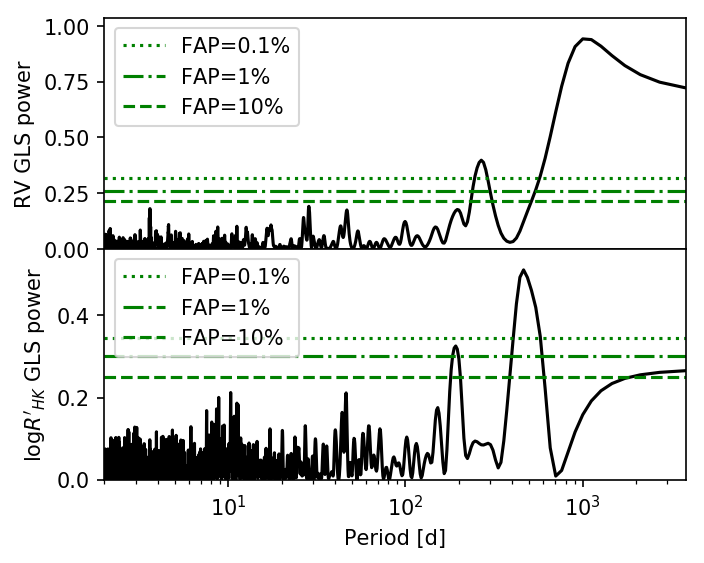}
 \caption{GLS periodograms of the spectroscopic time series. \textit{Upper panel:} RV. \textit{Lower panel:} $\log R'_\text{HK}$.}
 \label{fig:periodograms_rv_rhk}
\end{figure}

Fig. \ref{fig:periodograms_rv_rhk} shows the GLS periodograms of the RV and $\log R'_\text{HK}$. The RV periodogram identified as the strongest periodic signal at $f = 0.000999 \pm 0.000025$ d$^{-1}$ ($P = 1001 \pm 25$ d), with an amplitude $K = 266.1 \pm 7.0$ m s$^{-1}$, and a bootstrap false alarm probability, FAP$<10^{-4}$ \citep{endletal2001}. It is worth noticing that, as apparent from Fig. \ref{fig:rv_TERRA_Gaia}, the RV signal present in the time series is highly non-sinusoidal, and thus the sine-wave fitting performed by GLS is a poor match to the data. Nevertheless, the period shows good correspondence with the value from Gaia DR3 (see Table \ref{tab:gaia_sol}).

\begin{figure}
   \centering
   \includegraphics[width=.45\textwidth]{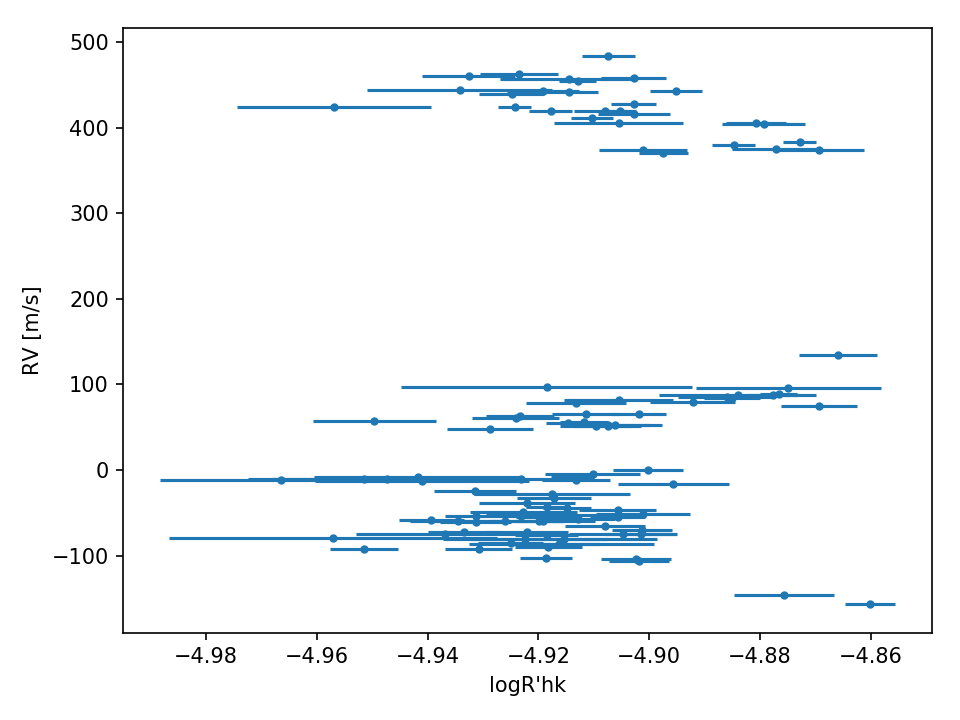}
 \caption{RV-activity correlation: RV time series as a function of $\log R'_\text{HK}$.}
 \label{fig:rv_rhk_corr}
\end{figure}

The amplitude of the signal is much larger than the value derived from the astrometric orbital parameters (i.e. $K \simeq 74$ m s$^{-1}$) as observed in Fig. \ref{fig:rv_TERRA_Gaia}. The $\log R'_\text{HK}$ periodogram shows a significant periodicity at $P = 460$ d. This however, does not appear to be connected with the RV signal, as the two time series show a very low Pearson (Spearman) correlation, $\rho = 0.21$ ($r_s = 0.25$), as shown in Fig. \ref{fig:rv_rhk_corr}. The only other significant peak in the $\log R'_\text{HK}$ periodogram is at $P \simeq 230$ d, close to half the first peak period. After removing these two signals, no other significant peak is found.

\begin{figure}
   \centering
   \includegraphics[width=.45\textwidth]{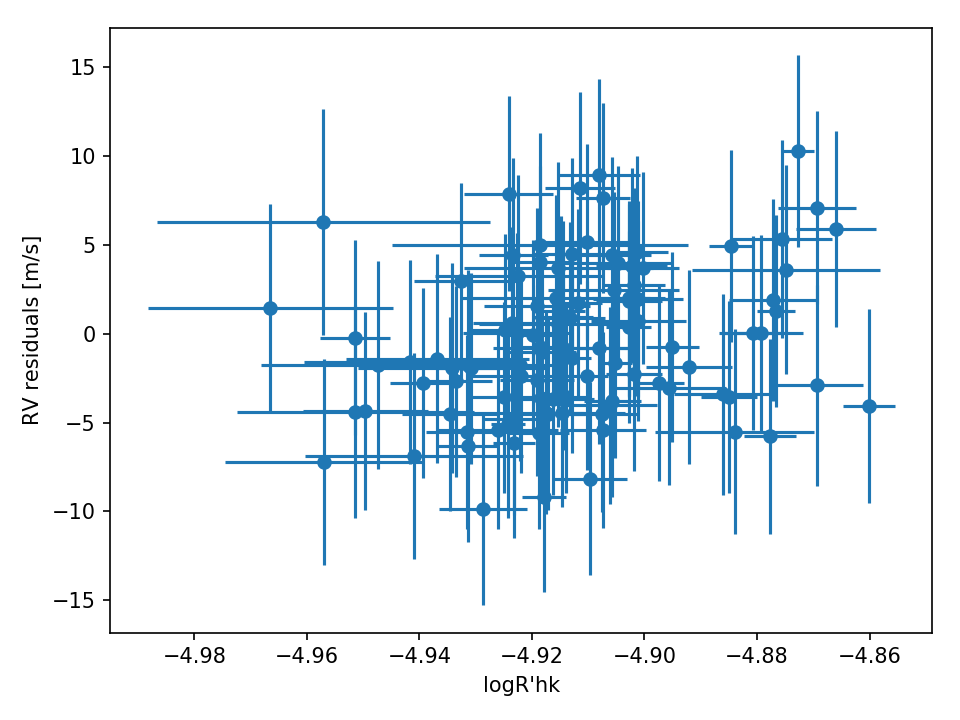}
 \caption{RV-activity correlation: RV residuals time series as a function of $\log R'_\text{HK}$.}
 \label{fig:rv_rhk_corr_res}
\end{figure}

As an additional test of the influence of stellar activity on the RV time series, we tested the correlation between the  $\log R'_\text{HK}$ and the residuals of the one-Keplerian model (see Sect. \ref{sec:1_plan}). The correlation appears to be only slightly higher than in the original time series,  $\rho = 0.26$ ($r_s = 0.29$), as shown in Fig. \ref{fig:rv_rhk_corr_res}. This again does present evidence for a strong stellar influence on the RV data.

\subsection{MCMC RV Keplerian fit}
\label{sec:1_plan}

We then fit a Keplerian orbit to the HARPS-N RV time series via the MCMC framework described above. We adopted uninformative uniform priors for the Keplerian parameters, semi-amplitude, $K_\text{b}$, orbital period, $P_\text{b}$, and time of periastron passage, $T_{p,b}$, as listed in Table \ref{tab:mcmc_prior_param}. For the eccentricity and argument of periastron, we followed the $\sqrt{e} \cos{\omega}$ and $\sqrt{e} \sin{\omega}$ parametrisation from \citet{eastman13}.
The RV model and residuals are shown in Fig. \ref{fig:one_plan_model}. From this model, we obtained an amplitude of $K_\text{b} = 321.9^{+3.6}_{-3.4}$ m s$^{-1}$, a period of $P_\text{b} = 2607^{+84}_{-79}$ d, and an eccentricity of $e_\text{b} = 0.8061^{+0.0064}_{-0.0065}$, corresponding to a minimum mass of $M_\text{b} \sin i = 14.67^{+0.50}_{-0.51}$ $M_\text{J}$. The complete posterior distributions of the model are shown in Fig. \ref{fig:one_plan_posterior}. These values are significantly different from the Gaia DR3 solution, with the longer period and higher eccentricity resulting in a larger mass, at the threshold between GPs and BDs.

\begin{figure}
   \centering
   \includegraphics[width=.45\textwidth]{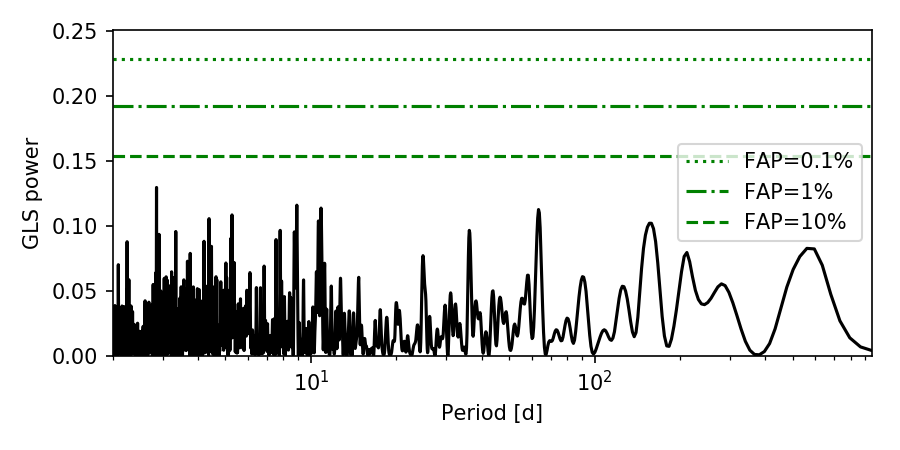}
 \caption{GLS periodogram of the RV residuals of the Keplerian model.}
 \label{fig:one_plan_peri_res}
\end{figure}

Two outliers can be identified in the RV residuals in Fig. \ref{fig:one_plan_model}, more than $3\sigma$ away from the median.
These data do not stand out in the $\log R'_\text{HK}$ time series shown in Fig. \ref{fig:rv_TERRA_Gaia}, so we checked additional indicators to identify the source of these outliers. Checking other CCF activity indicators, the second epoch (BJD=2459844.32) stands out as an outlier in the FWHM and BIS time series (see Appendix \ref{app:activity}), while the first  (BJD=2459830.33) was taken during a night of bad weather and very variable seeing. Thus, we decided to remove both these epochs in all the following analyses to avoid contaminations in the subsequent tests.
We then performed a GLS periodogram on the RV residuals to test the presence of additional signal, planetary or stellar, in the time series. As shown in Fig. \ref{fig:one_plan_peri_res}, no significant signal was found.

\subsection{Additional RV models}
\label{sec:add_rv}

Although no significant peak was found in the GLS periodogram of the residuals, we tested a few additional MCMC models to investigate in depth the presence of potential additional signals in the data. We also tested an additional fit to see if the Gaia DR3 orbital solution could be reconciled with the results of our RV Keplerian model. 

\subsubsection{Long-term trend model}
We also tested the presence of additional long-period signals in the data in the form of a linear acceleration term added to the model, $d \cdot (t-\bar{t})$. As shown in Table \ref{tab:mcmc_prior_param}, this model shows a worse BIC than the previous one without the linear trend ($\Delta\text{BIC} = +5$). Moreover, the best-fit value for the acceleration term is not significantly different from zero. For these reasons, we decided to discard this model and adopt as final model the previous one.

\subsubsection{Two-Keplerian model}
As an additional test on the presence of short-period companions in the system, we run an MCMC two-Keplerian model, with wide priors for the second Keplerian period, $P_c \in \mathcal{U}$(1.5,100) d (model not shown). The MCMC did not converge to any significant periodicity, and the two-Keplerian model is statistically disfavoured with respect to the one-Keplerian one ($\Delta\text{BIC} = +5$).

\begin{figure}
   \centering
   \includegraphics[width=.45\textwidth]{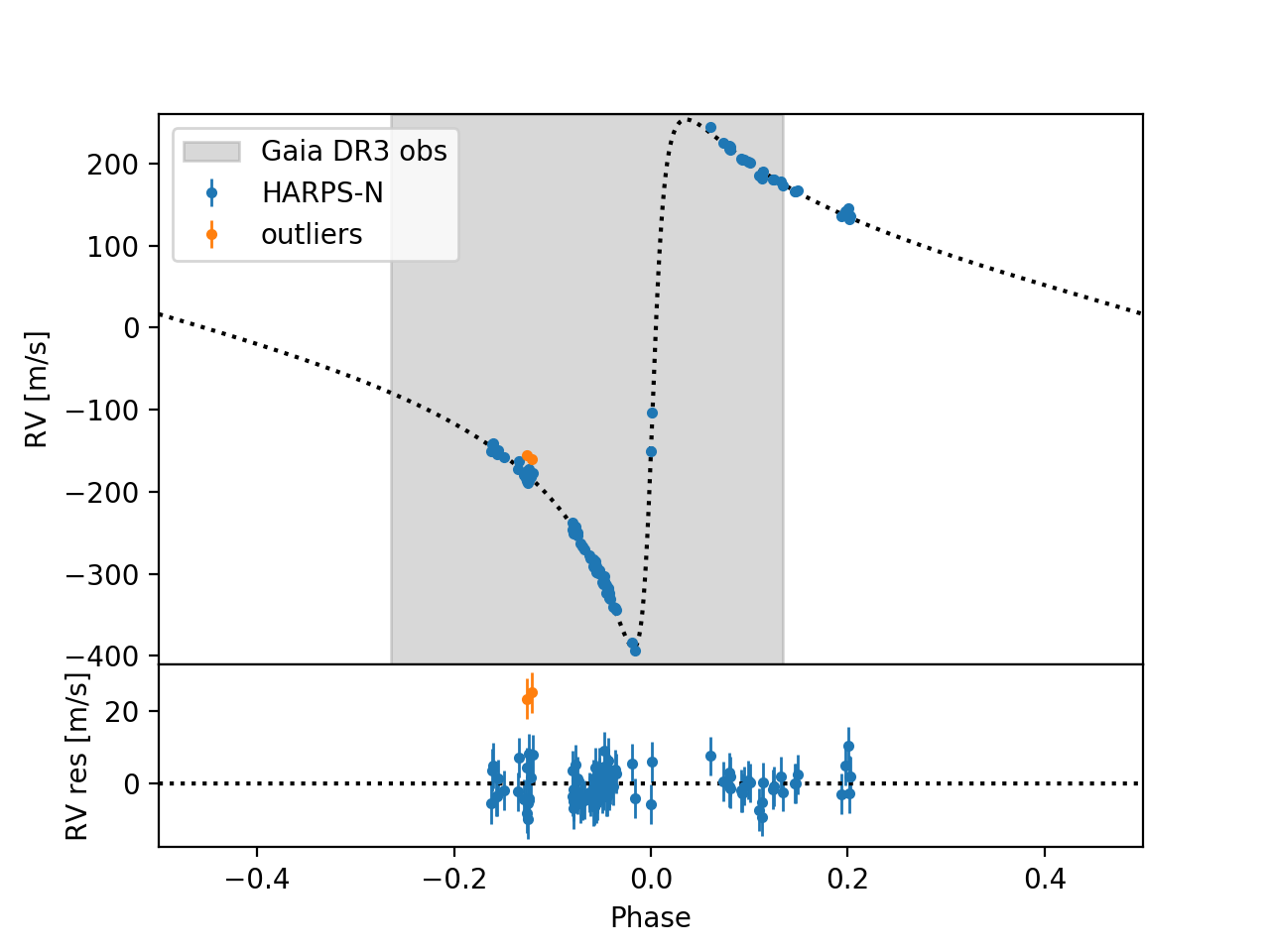}
 \caption{Phase-folded best-fit RV one-Keplerian model, corrected for the instrumental offset (top panel) and the RV residuals (bottom panel). The orange dots represents the outliers discussed in Sect. \ref{sec:1_plan}, while the grey-shaded area represent the phase coverage of Gaia DR3 observations.}
 \label{fig:one_plan_model}
\end{figure}

\begin{table*}
\caption[]{Priors and best-fit parameters for the MCMC models tested in Sect. \ref{sec:kep_fit}.}
\label{tab:mcmc_prior_param}
\centering
\begin{tabular}{lccc}
\hline
\hline
\noalign{\smallskip}
 & \multicolumn{2}{c}{RV 1-Keplerian} & RV 1-Keplerian + trend \\ 
& Priors   & Values & Values  \\
\hline
\noalign{\smallskip} 
$K_\text{b}$ $[$m s$^{-1}]$ & $\mathcal{U}$(0,1000) & $321.9^{+3.6}_{-3.4}$ & $321.2^{+3.8}_{-3.6}$  \\ 
\noalign{\smallskip}  
$P_\text{b}$ $[$d$]$ & $\mathcal{U}$(500,3000) & $2607^{+84}_{-79}$ & $2423^{+365}_{-34}$  \\ 
\noalign{\smallskip}  
$T_{p,b}$ $[$BJD$-2450000]$ & $\mathcal{U}$(9700,10500) & $10158.73^{+0.59}_{-0.60}$ & $10158.83^{+0.60}_{-0.61}$ \\
\noalign{\smallskip}  
$\sqrt{e_\text{b}} \cos{\omega_\text{b}}$ & $\mathcal{U}$(-1,1) & $-0.236^{+0.011}_{-0.011}$ & $-0.232^{+0.012}_{-0.012}$ \\
\noalign{\smallskip}
$\sqrt{e_\text{b}} \sin{\omega_\text{b}}$ & $\mathcal{U}$(-1,1) & $-0.8660^{+0.0048}_{-0.0047}$ & $-0.861^{+0.011}_{-0.009}$  \\
\noalign{\smallskip}
\hline
\noalign{\smallskip}  
$M_\text{b} \sin i$ $[M_\text{J}]$  &  & $14.67^{+0.50}_{-0.51}$ & $14.59^{+0.52}_{-0.52}$  \\ 
\noalign{\smallskip}  
$a_\text{b}$ $[$AU$]$  &  & $3.95^{+0.11}_{0.10}$ & $3.76^{+0.37}_{-0.35}$  \\ 
\noalign{\smallskip}  
$e_\text{b}$  &  & $0.8061^{+0.0064}_{-0.0065}$  & $0.796^{+0.018}_{-0.020}$   \\ 
\noalign{\smallskip}  
$\omega_{\star,b}$ $[$rad$]$ &  & $-1.837^{+0.013}_{-0.013}$ & $-1.834^{+0.013}_{-0.013}$  \\ 
\noalign{\smallskip}  
\hline  
\noalign{\smallskip}  
$\gamma_\text{HARPS-N}$ $[$m s$^{-1}]$ & $\mathcal{U}$(-50.0,250.0) & $238.14^{+1.01}_{-0.99}$ & $237.2^{+1.9}_{-1.9}$  \\ 
\noalign{\smallskip}  
$\sigma_\text{jit,HARPS-N}$ $[$m s$^{-1}]$ & $\mathcal{U}$(0,100) & $5.22^{0.44}_{-0.39}$ & $5.22^{+0.44}_{-0.38}$  \\ 
\noalign{\smallskip} 
$d$ $[$m s$^{-1}$ d$^{-1}]$ & $\mathcal{U}$(-0.05,0.05) & - & $0.010^{+0.024}_{-0.019}$  \\ 
\noalign{\smallskip} 
\hline  
\noalign{\smallskip}  
BIC & & $688$ & $693$  \\ 
\hline
\end{tabular}
\tablefoot{The best RV model is highlighted in bold.}
\end{table*}

\section{Gaia simulations}
\label{sec:gaia_sim}

To further study the discrepancy between the RV and astrometric solutions, we performed a suite of numerical simulations. First, we produced a set of realistic Gaia astrometric time series, injecting possible realisations of a Keplerian signal corresponding to the RV orbital fit. We then studied the orbital solutions recovered by fitting these synthetic astrometric data. This process  was meant to help us explain why the Gaia DR3 solution was so different from our findings and to further study whether Gaia DR4 could help us confirm our results. 

The setup of this simulation is the same as in \citet{sozzettietal2023} and \citet{ruggierietal2024}. We used GOST\footnote{Gaia Observation Forecast Tool - \url{https://gaia.esac.esa.int/gost/index.jsp}} to obtain the Gaia observation information (observation time, scan angle, and parallax factor) and took the five-parameter stellar motion solution from Gaia DR3. We took the RV orbital parameters
from our HARPS-N solution, drawing them from a Gaussian distribution. We generated the longitude of the ascending node, $\Omega$, drawn from a uniform distribution, $[0, \pi]$, and the inclination, $i$, drawn from a uniform distribution of $\cos i \in [-1, 1]$.

\subsection{DR3 simulations}
\label{sec:dr3_sim}

We simulated the Gaia DR3 time series obtaining from GOST all the Gaia observations of HD\,128717 from July 2014 to May 2017 \citep{gaiaetal2021}. As measurement error, we adopted $\sigma = 150$ $\mu$as, which is representative of the DR3 astrometric uncertainties for a star of magnitude G $=8.2$ \citep{holletal2023}.

\begin{figure}
   \centering
\includegraphics[width=.45\textwidth]{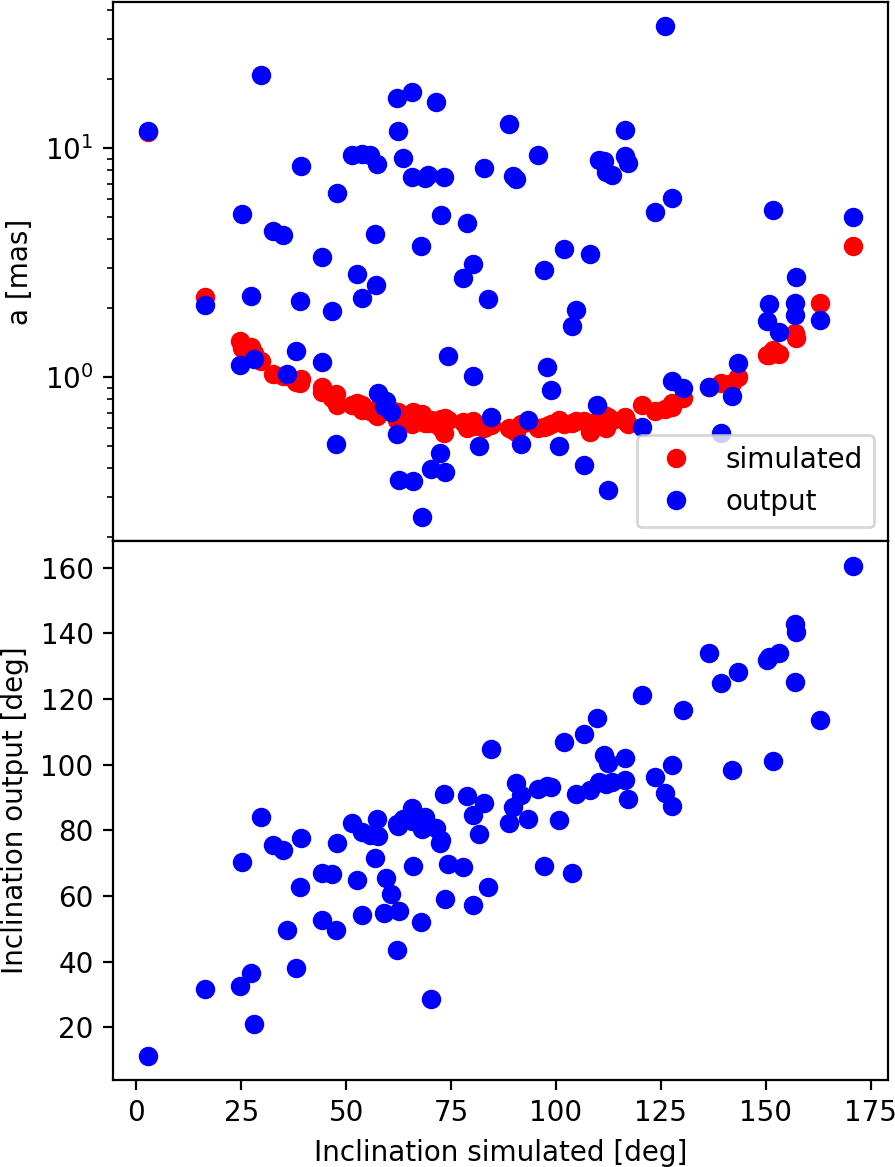}
 \caption{Astrometric simulation results for DR3. \textit{Upper panel:} Simulated (red) and derived (blue) astrometric semimajor axis vs injected orbital inclination; \textit{Lower panel:} Derived inclination vs the simulated value.}
 \label{fig:dr3_orb_corr}
\end{figure}

\begin{figure}
   \centering
   \includegraphics[width=.45\textwidth]{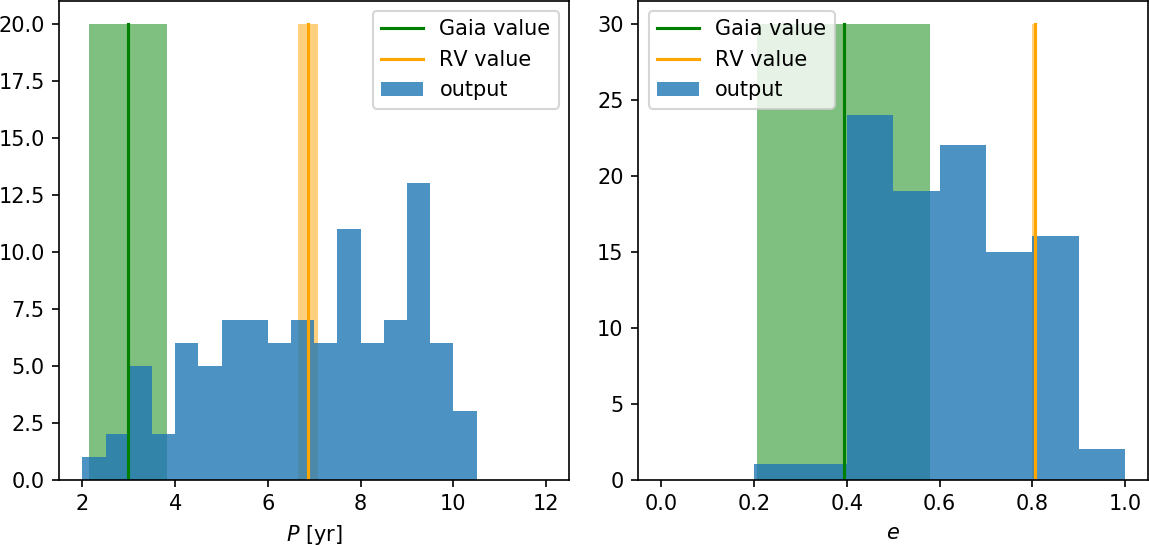}
 \caption{Distribution of the recovered values of period, $P$, on the left, and eccentricity, $e$, on the right, in the DR3 astrometric simulations. The green lines and shaded regions mark the best-fit values and uncertainties from the Gaia DR3 orbital solution, while the orange lines and regions represent our RV solution.}
 \label{fig:dr3_sim_dist}
\end{figure}

We produced 100 realisations of the astrometric time series and fit each one with a partly linearised model \citep{holletal2023} implemented with \texttt{emcee}. The best-fit values of the inclination, $i$, and astrometric semi-major axis, $a_0$, compared with the injected values, are shown in Fig. \ref{fig:dr3_orb_corr}. We can see that although there is a partial correspondence between the injected and recovered values of $i$ (lower panel Fig. \ref{fig:dr3_orb_corr}), the semi-major axis, $a_0$, is almost never recovered correctly (see upper panel Fig. \ref{fig:dr3_orb_corr}). This  also implies  that the orbital period is very hard to constrain from the DR3 astrometric data. This is highlighted in Fig. \ref{fig:dr3_sim_dist}, where the distributions of period and eccentricity are compared with the values from the RV and Gaia DR3 orbital solutions. We can see that the distributions do not resemble the RV values, although they were used to generate the synthetic time series and are very wide. In particular, the eccentricity peaks at lower values, close to the value of $e = 0.39$ obtained by Gaia DR3. If we compute how many times the fitted value is within 1$\sigma$ from the RV value of $P$ and $e$, we get six and three times out of a total of 100 simulations, respectively. For the Gaia DR3 solution, we fall within 1$\sigma$ 10 and 42 times, respectively. Thus, we can see that even if there is no clear bias towards the Gaia DR3 orbital solution, it can be easily recovered from the astrometric time series thanks to the very wide distribution of the possible solutions. This is not unexpected, since the RV data clearly indicate an orbital period much longer than the duration of DR3 observations and, thus, the signal is very difficult to constrain due to the poor orbital coverage.

\subsection{DR4 simulations}
\label{sec:dr4_sim}

\begin{figure}
   \centering \includegraphics[width=.45\textwidth]{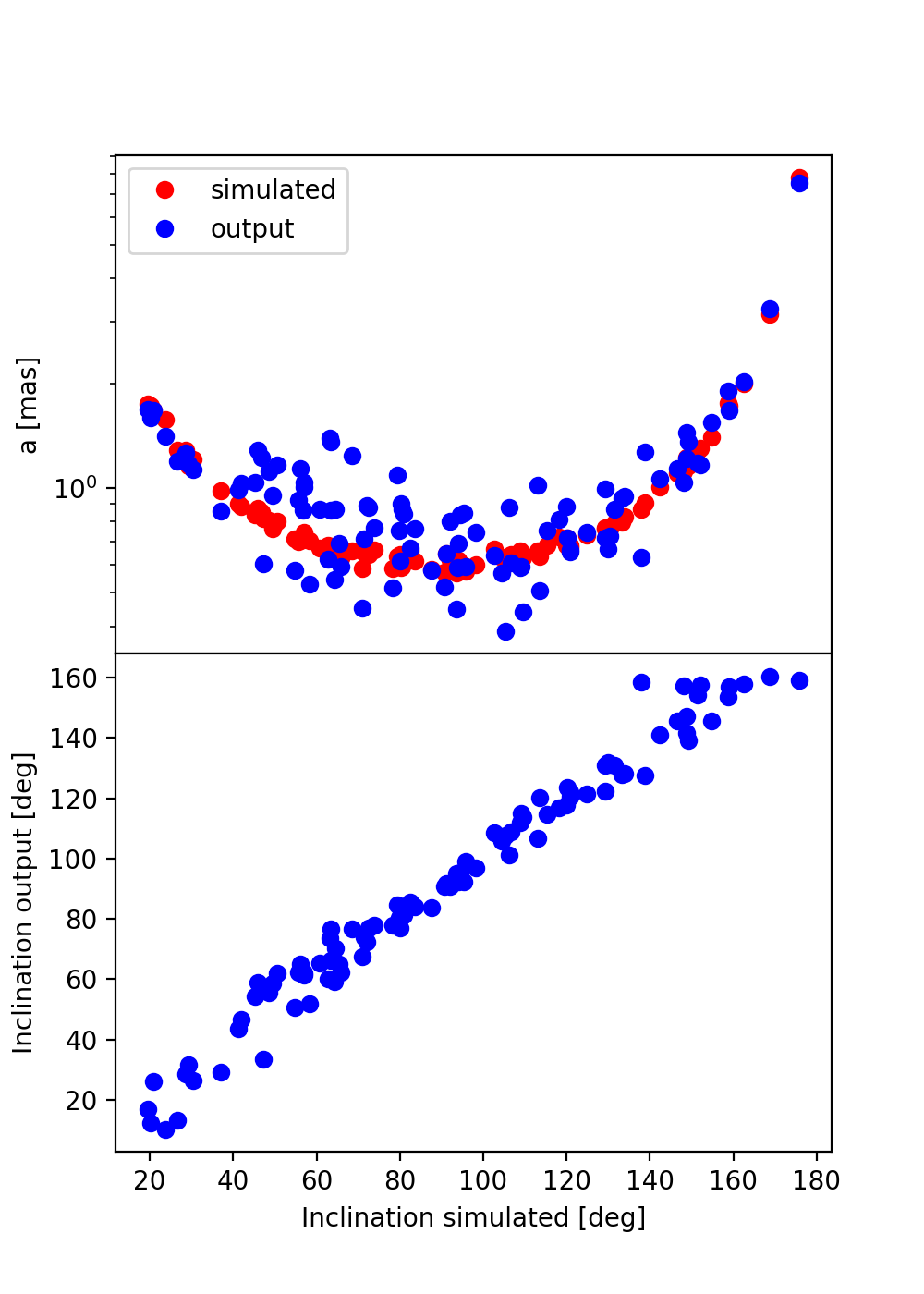}
 \caption{Astrometric simulation results for DR4. \textit{Upper panel:} Simulated (red) and derived (blue) astrometric semimajor axis vs injected orbital inclination; \textit{Lower panel:} Derived inclination vs simulated value.}
 \label{fig:dr4_orb_corr}
\end{figure}

\begin{figure}
   \centering
   \includegraphics[width=.45\textwidth]{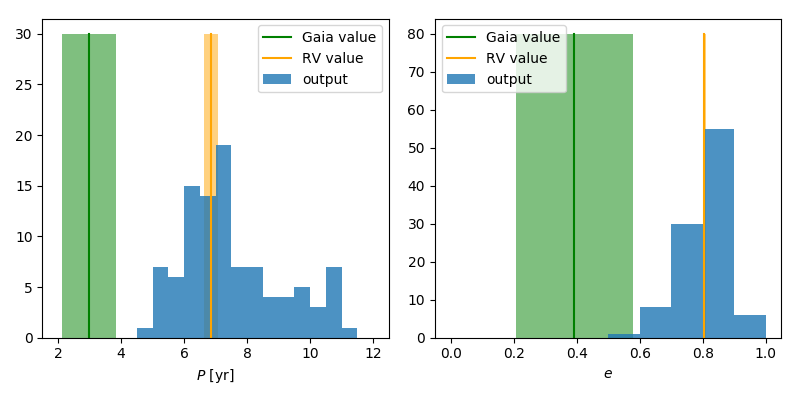}
 \caption{Distribution of the recovered values of period, $P$, and eccentricity, $e$, in the DR4 astrometric simulations. The green lines and shaded regions mark the best-fit values and uncertainties from the Gaia DR3 orbital solution, while the orange lines and regions represent our RV solution.}
 \label{fig:dr4_sim_dist}
\end{figure}

Next, we  produced a set of 100 Gaia DR4 time series with the GOST observations from July 2014 to December 2019. We adopted internal uncertainties of 80 $\mu$as, which is expected to be a representative value for similar stars in DR4 \citep{brown2024}. We can see from Fig. \ref{fig:dr4_orb_corr} that the correspondence between the injected and recovered values is much better with the DR4 time series, with respect to the DR3 simulations discussed previously.
Although some uncertainty on the semi-major axis remains, the orbital inclination is recovered with great accuracy. In Fig. \ref{fig:dr4_sim_dist}, we can see how the distribution of the output values of $P$ and $e$ closely peaks to the RV orbital solution used to produce the datasets.
These results show how the current DR3 time span is insufficient to correctly recover the Keplerian orbit, while the next data release is expected to have sufficient coverage to accurately recover the companion signal.

\section{Combined fit of RVs and absolute astrometry}
\label{sec:pma_fit}

\begin{table}[ht!]
    \centering
       \small
        \caption{Orbital parameters and true mass of HD\,128717\,B from a combined RV+absolute astrometry analysis. \label{tab:fit_HD128717}
}
        \begin{tabular}{lcc}
    \hline
    \hline
    \noalign{\smallskip}
    Parameter     &  Prior &  Value \\
    \noalign{\smallskip}
    \hline
    \noalign{\smallskip}
    \noalign{\smallskip}
    $P_\text{b}$ [yr] & $\mathcal{U}(5.0,25.0)$  & $9.37^{+0.06}_{-0.05}$  \\
    \noalign{\smallskip}
    $T_{p,b}$ [yr] & $\mathcal{U}(0.0,3000.0)$  & $2023.6716^{+0.0009}_{-0.0009}$  \\
    \noalign{\smallskip}
    $a_\text{b}$ [au] & $\mathcal{U}(1.0,10.0)$  & $4.85^{+0.02}_{-0.01}$  \\
    \noalign{\smallskip}
    $e_\text{b}$  & $\mathcal{U}(0.0,1.0)$  & $0.850^{+0.002}_{-0.002}$  \\
    \noalign{\smallskip}
    $\omega_\text{b}$ [deg]  & $\mathcal{U}(0.0,360.0)$  & $-105.7^{+0.4}_{-0.3}$ \\
    \noalign{\smallskip}
    $i_\text{b}$ [deg] & $\cos(i):\,\mathcal{U}(-1.0,1.0)$ &  $130.3^{+1.6}_{-1.9}$ \\
    \noalign{\smallskip}
    $\Omega_\text{b}$ [deg] & $\mathcal{U}(-180.0,180.0)$  & $131.7^{+3.3}_{-3.4}$ \\
    \noalign{\smallskip}
    $q_\text{b}$  & $\mathcal{U}(0.0,0.1)$  & $0.0148^{+0.0004}_{-0.0004}$  \\
    \noalign{\smallskip}
    $M_\text{b}$ $[M_\text{J}]$ & (derived) & $19.8^{+0.5}_{-0.5}$ \\
    \noalign{\smallskip}
    \noalign{\smallskip}
    \hline
    \end{tabular}
\end{table}

As reported in cross-calibrated Hipparcos-\textit{Gaia} catalogues of astrometric accelerations \citep{Brandt2021,kervellaetal2022}, HD\,128717 is clearly identified as an accelerating star, with a S/N of the proper motion anomaly (PMa) of $\sim9$ at the mean epoch of Gaia DR3. We performed a combined PMa+RV model to the Hipparcos-\textit{Gaia} absolute astrometry of HD\,128717 from \citet{Brandt2021} and to the offset-corrected RV dataset in order to determine the possible values of orbital inclination, $i$, longitude of the ascending node, $\Omega$, and mass ratio, $q$, which can be directly constrained by the PMa astrometric data. We adopted uninformative priors for the model parameters derived by the RV-only analysis and uniform priors on $\cos(i)$, $\Omega$, and $q$. 

The combined analysis utilises a differential evolution MCMC (DE-MCMC) algorithm \citep{terbraak2006,eastman13}, with the PMa model constructed by averaging over the actual observing windows of Hipparcos (utilising the exact time of observation of the target; see \citealt{VanLeeuwen2007}) and \textit{Gaia} DR3 (based on the accurate representation of the actual \textit{Gaia} transit times from GOST. 
The DE-MCMC analysis was run with standard prescriptions on the burn-in phase and convergence conditions based on the Gelman-Rubin statistics (e.g. \citealt{eastman13}). The medians of the posterior distributions of the model parameters constituted the nominal best-fit values,  with $1\sigma$ uncertainties determined evaluating the $\pm34$th percentile intervals of the posteriors (for additional details, see \citealt{sozzetti2023}). 

\begin{figure}[h!]
    \centering
    \includegraphics[width=0.45\textwidth]{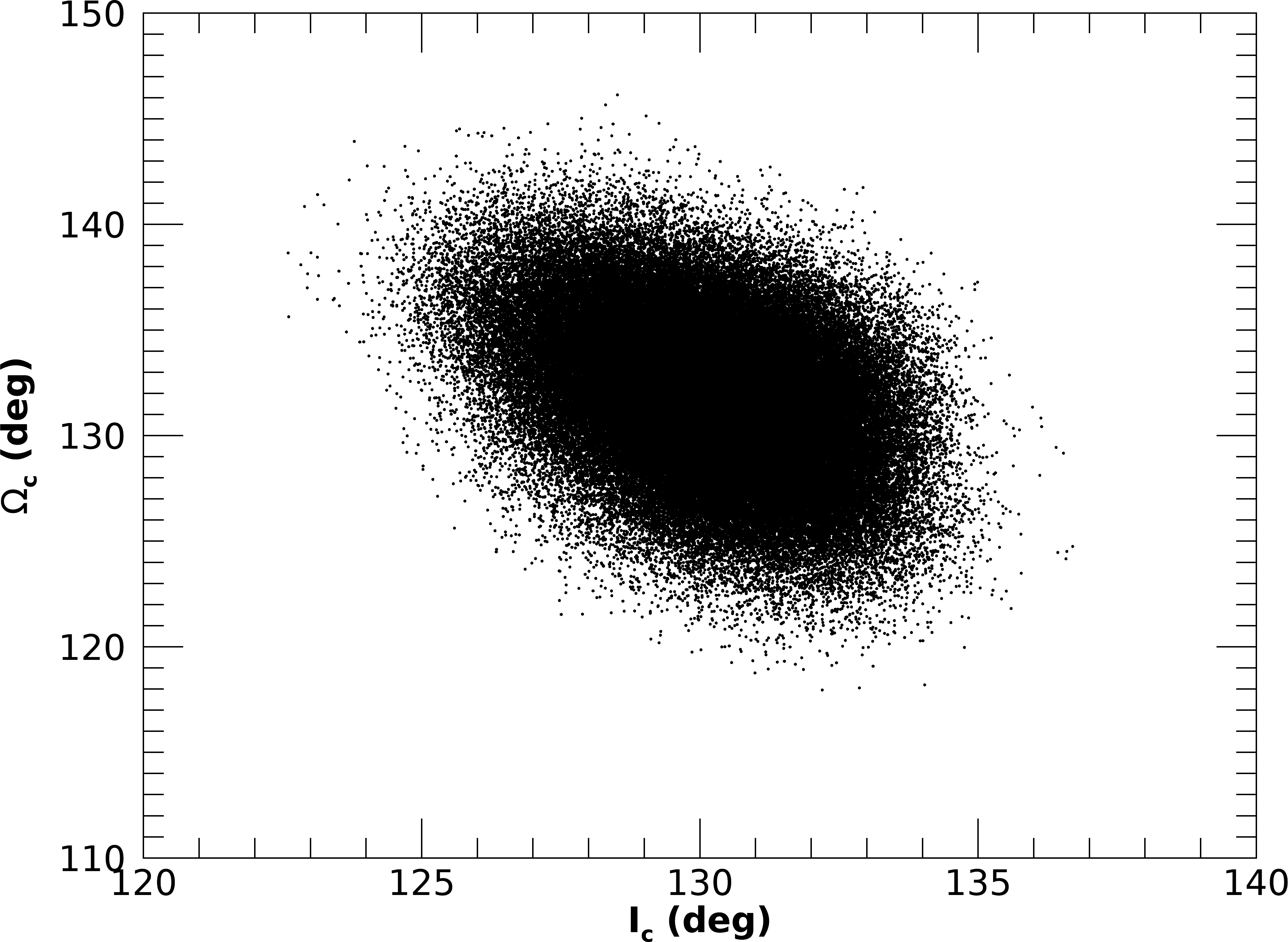} 
    
    \includegraphics[width=0.45\textwidth]{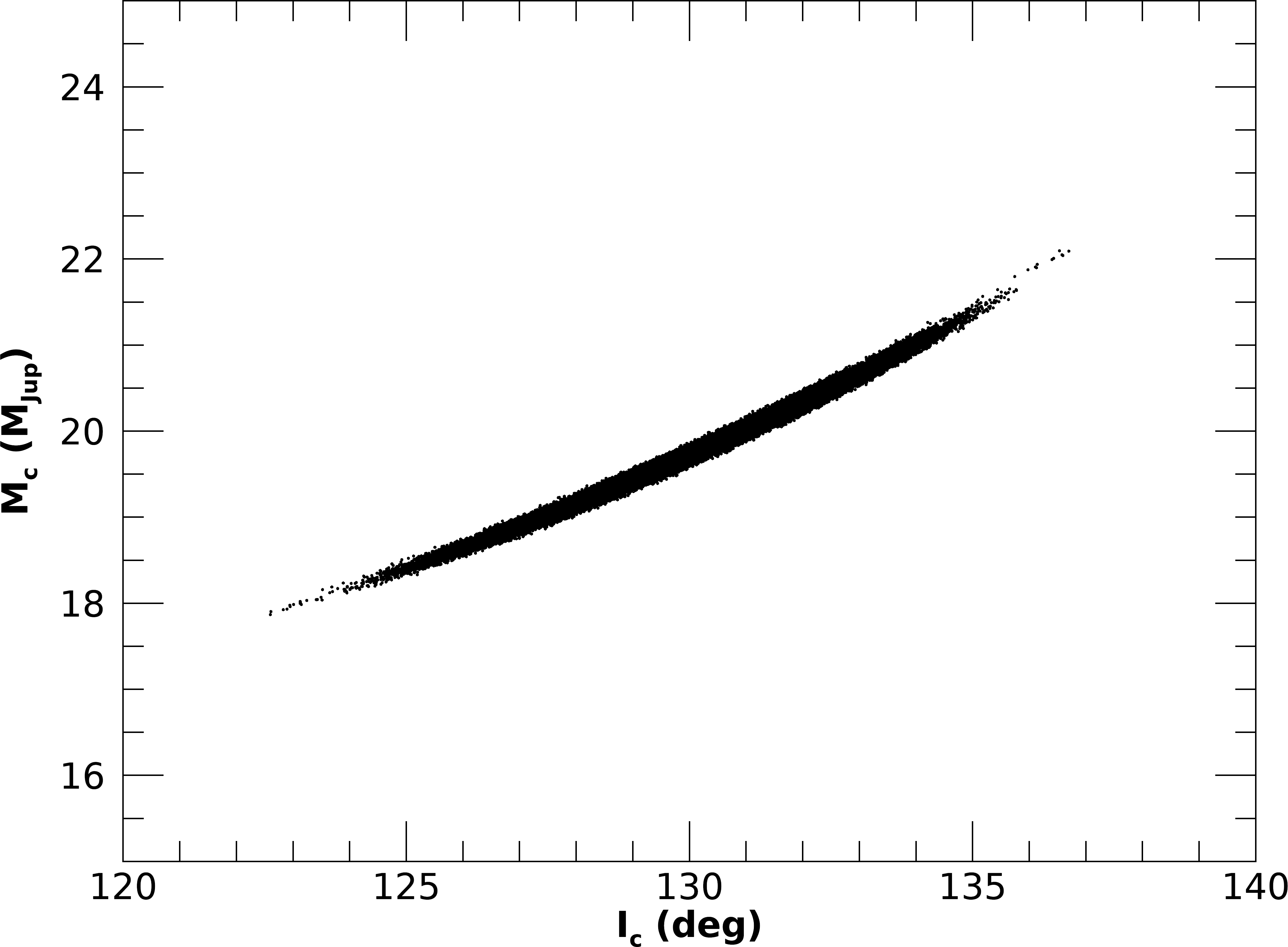} 
    \caption{Joint posterior distributions for inclination, longitude of the ascending node and true mass for HD\,128717\,B.}
    \label{fig:posteriors_HD128717}
\end{figure}

\begin{figure}
    \centering
    \includegraphics[width=0.45\textwidth]{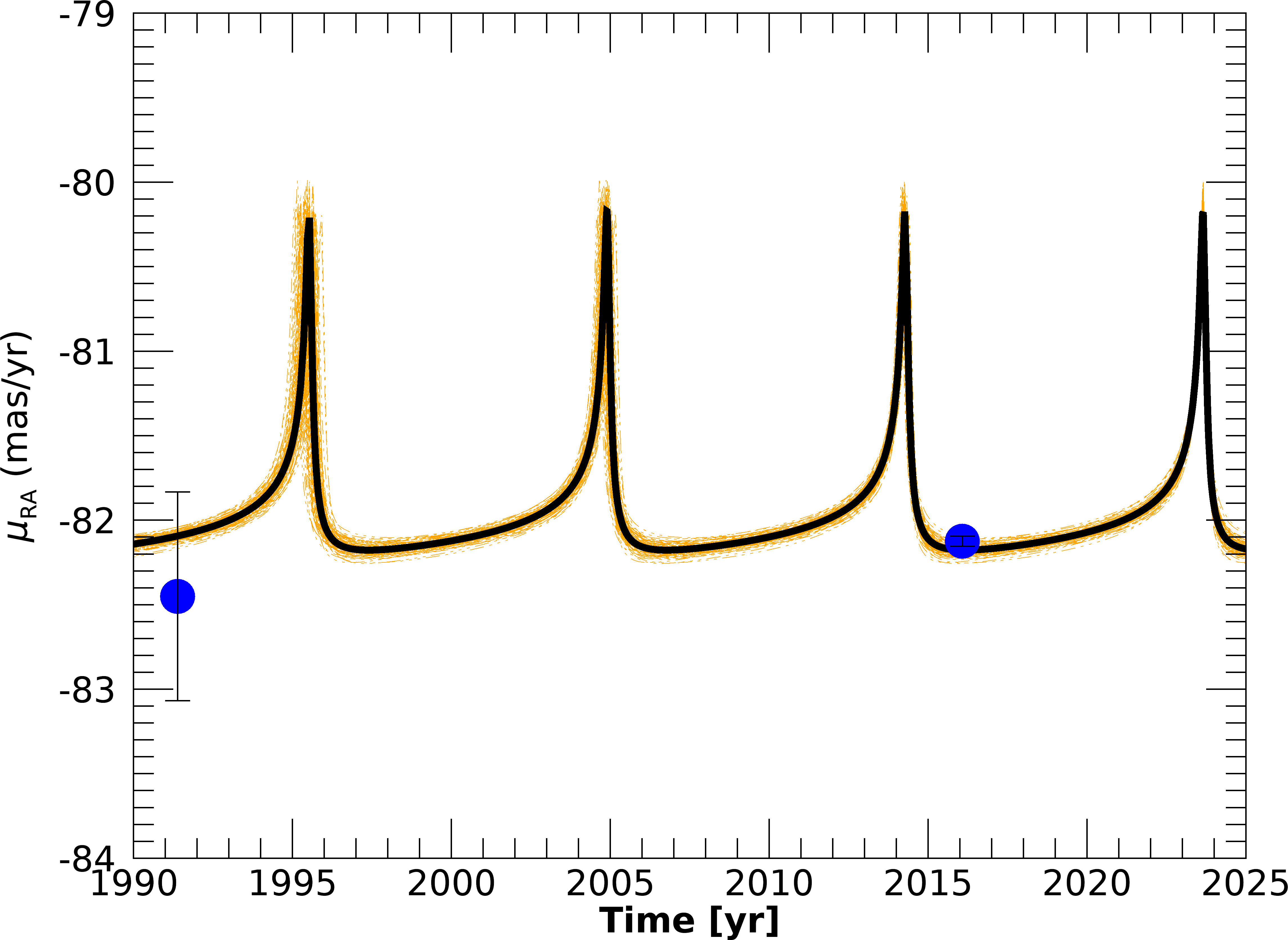} 
    
    \includegraphics[width=0.45\textwidth]{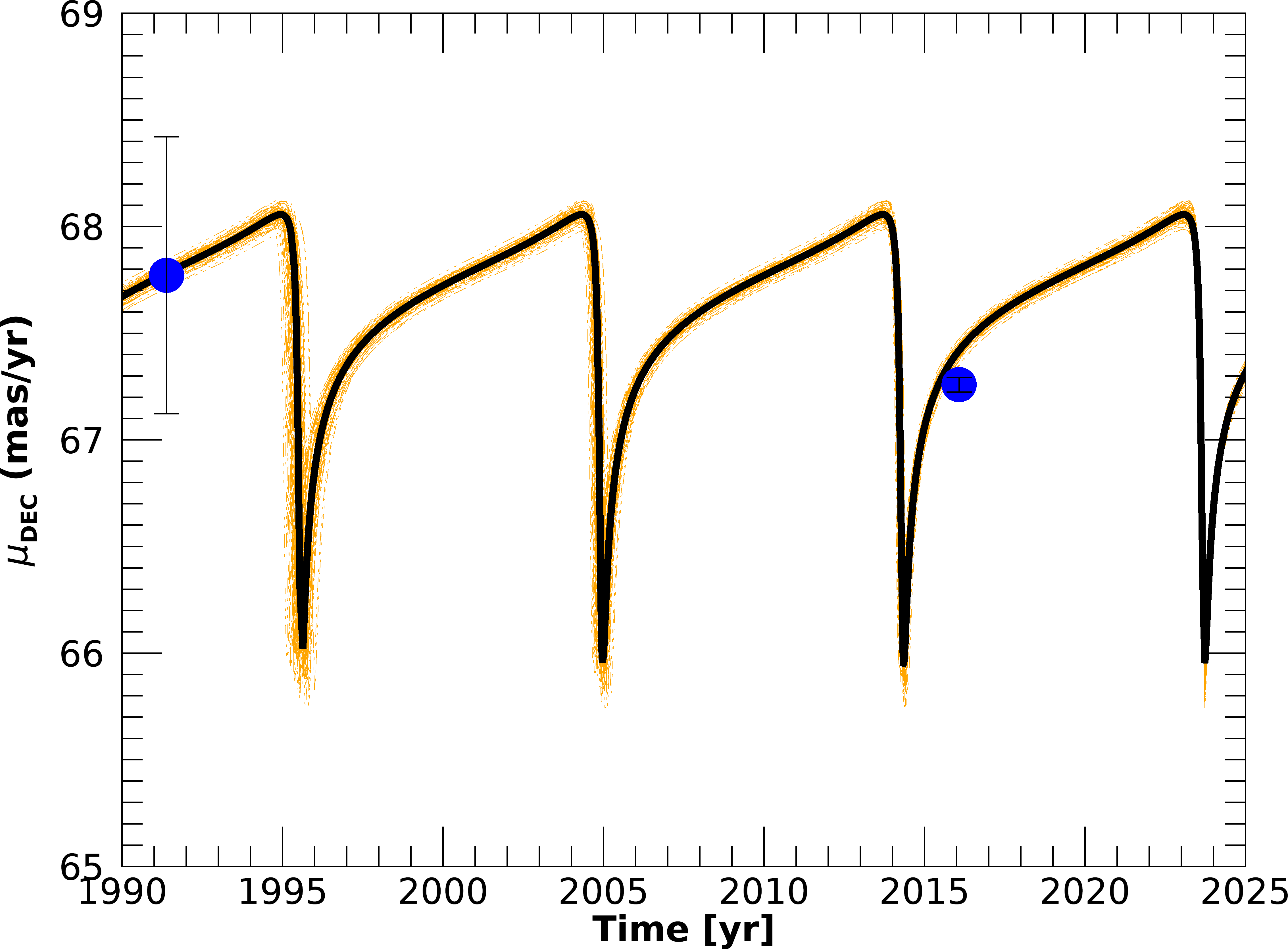} 
    
    \includegraphics[width=0.45\textwidth]{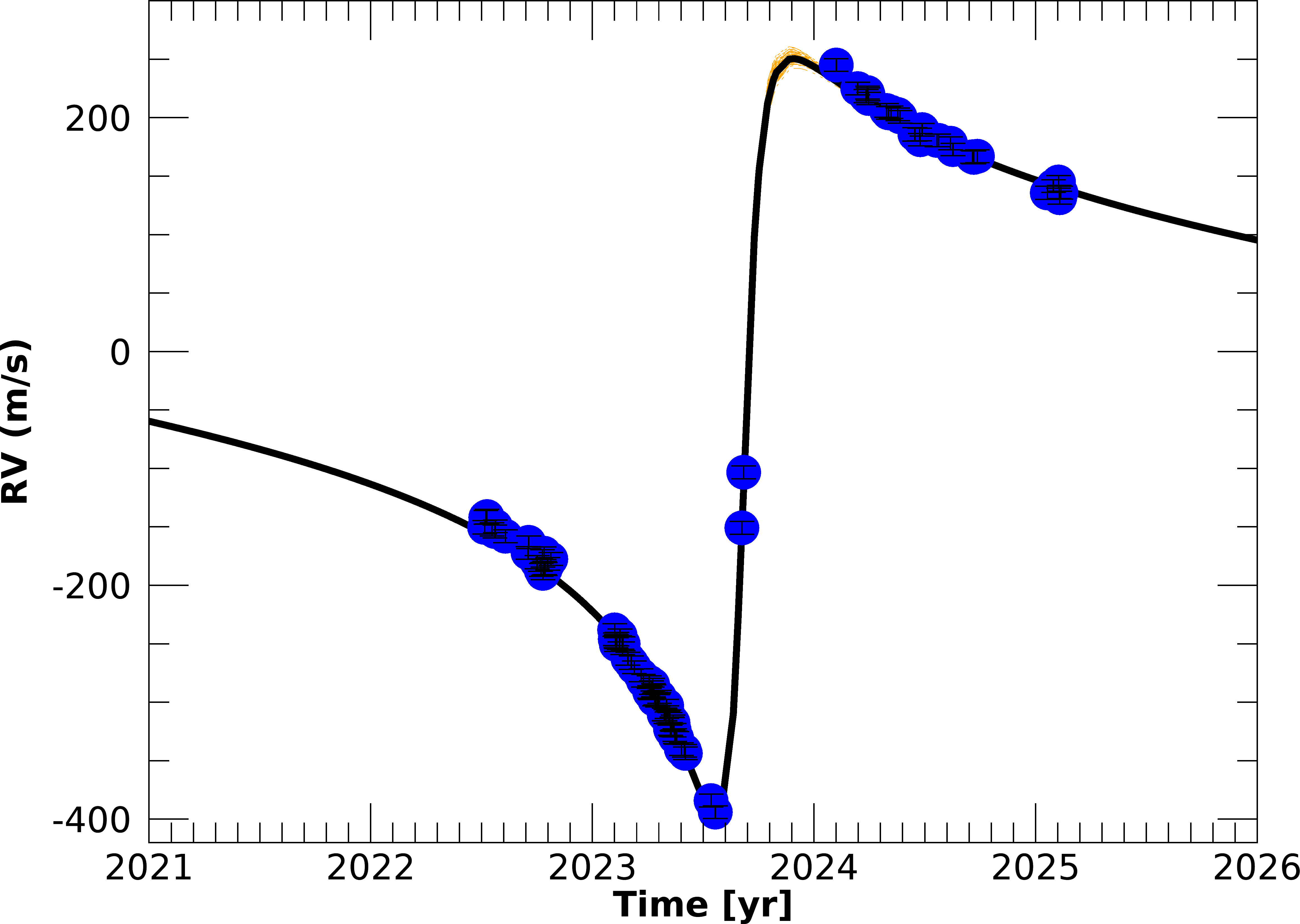}
    \caption{Best-fit orbital solution (solid black line) for HD\,128717\,B, superposed to the calibrated proper motions from Hipparcos and \textit{Gaia} (top and middle panels) and to the RV time series (bottom panel). The dashed orange lines represent random selections of orbital solutions drawn from the posterior distributions. }
    \label{fig:orbit_HD128717}
\end{figure}

Table \ref{tab:fit_HD128717} reports the best-fit orbital solution for HD\,128717\,B. Usually, applications of the PMa technique (e.g. \citealt{lietal2021,sozzetti2023}) return bimodal distributions of the inclination angle, due to the difficulty in separating prograde from retrograde orbits. In this case, instead of being bimodal, the inclination distribution has a single value of $i_\text{b}=130.3^{+1.6}_{-1.9}$ deg (with a corresponding $\Omega_\text{b}=131.7^{+3.3}_{-3.4}$ deg), possibly because the time interval encompassed by the Gaia DR3 observations critically samples the high-velocity phase of the very eccentric orbit of HD\,128717\,B just after periastron passage (see Fig. \ref{fig:one_plan_model}).
We obtained a mass ratio $q_\text{b}=0.0148\pm0.0004$ and derive a true mass of $M_\text{b}=19.8\pm0.5$ $M_\text{J}$. The combined solution indicates that HD\,128717\,B's orbit has a period about $30\%$ longer and a slightly higher eccentricity than that inferred by the RV-only fit. In Fig. \ref{fig:posteriors_HD128717} and Fig. \ref{fig:orbit_HD128717}, we show the joint posteriors of these parameters and the best-fit orbital solution superposed to the observational data, respectively. 

\section{Discussion}
\label{sec:discussion}

In light of the aforementioned analyses, we confirmed the Keplerian nature of Gaia-ASOI-009 b. It  is a sub-stellar companion of mass $M_\text{b} = 19.8\pm 0.5$ $M_\text{J}$ (hereafter, Gaia-6\,B).

\subsection{Comparison with Gaia DR3 solution}
Thanks to high-precision RV follow-up with the HARPS-N spectrograph of HD\,128717, we were able to confirm the presence and sub-stellar nature of the candidate Gaia-ASOI-009\,b. Nonetheless, our final Keplerian solution for Gaia-6\,B (see Table \ref{tab:fit_HD128717}) is significantly different from the one from Gaia DR3 (Table \ref{tab:gaia_sol}). An additional test to see whether the two solutions could be reconciled can be found in Appendix \ref{app:gaia_emcee} but, as expected, the evidence from the RV time series strongly points away from the Gaia DR3 solution.

This discrepancy can be explained using the simulations we performed, as discussed in Sect. \ref{sec:gaia_sim}. As an additional test, we produced 100 Gaia DR3 time series using the orbital parameters from the final PMa+RV solution (Table \ref{tab:fit_HD128717}) and compared the likelihood of the simulated data given two different set of parameters from the Keplerian model: the set derived from our PMa+RV fit and the one from the Gaia DR3 solution. We find that the two distributions of likelihood are indistinguishable: a Kolmogorov-Smirnov (K-S) test on the likelihood distributions returned a p-value $p=96\%$. Moreover, we see that in 42/100 cases the likelihood of the DR3 solution is higher than that of our final solution. This proves that the Gaia DR3 time series is not able to constrain the orbit of Gaia-6\,B. The reason for this is most probably the orbital period of the sub-stellar companion: the time span of Gaia DR3 observations (i.e. 34 months) is much shorter than the orbital period, $P_\text{b} = 9.37$ yr. Combined with the high eccentricity of the orbit, this results in a strong degeneracy between eccentricity and orbital period, which most likely produced the Gaia DR3 solution.

\begin{figure*}
    \centering
    \includegraphics[width=0.98\textwidth]{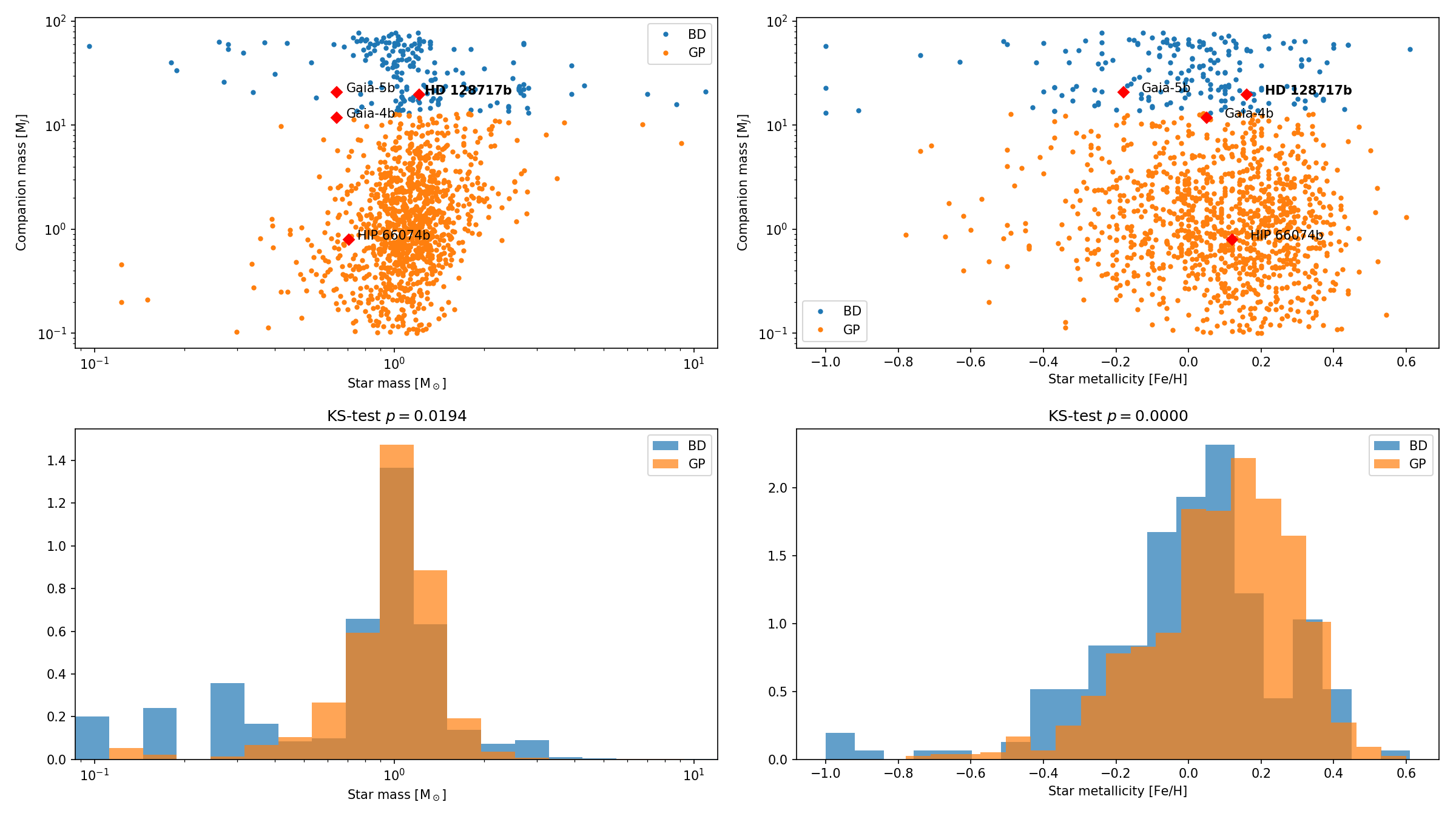}
    \caption{Masses of GPs and BDs as a function of stellar host mass and metallicity, for sub-stellar companions with period $<10^4$ d. BDs ($M>13$ $M_\text{J}$) are shown as blue points, while GPs ($0.1 <M<13$ $M_\text{J}$) are shown as orange points. The red diamonds highlight the position of confirmed Gaia-ASOI sub-stellar companions \citep[and HIP 66074\,B][]{sozzettietal2023}. The lower plots show the stellar mass and metallicity distributions of BD and GP. }
    \label{fig:brown_dwarfs}
\end{figure*}

\subsection{Giant planet-brown dwarf classification}

Given its high mass, $M_\text{b} = 19.8\pm 0.5$ $M_\text{J}$, the sub-stellar companion identified in the astrometric and RV data lies above the commonly adopted 13 $M_\text{J}$ threshold dividing `planets' and `BDs'; however, the exact threshold is somewhat arbitrary \citep[e.g.][and references therein]{stefanssonetal2025}.

In Fig. \ref{fig:brown_dwarfs}, we show a comparison of Gaia-6\,B with the known population of GPs and BDs, as defined in Appendix \ref{app:bd_gp_pop}. We can see that the mass of the host star is very close to the peak of both the BD and GP distributions. On the other hand, it has a higher-than-average metallicity, close to the peak of the GP distributions. This can lead to some uncertainty on the nature of the sub-stellar companions, depending on the definition adopted to distinguish planets from BDs.
Metallicity is commonly seen as a statistical indicator to distinguish between GPs and BDs \citep[e.g.][]{mage2014,matasanchezetal2014,maldonadovillaver2017}, as it can be correlated with the formation mechanisms: GPs formed via core accretion are more common around metal-rich stars \citep[e.g.][]{osbornbayliss2020}, while BDs formed via gravitational instability are considered to be independent of metallicity \citep[e.g.][]{maldonadoetal2019}. From this point of view Gaia-6\,B might belong to both populations, being located at the joint between the two populations, but its high metallicity could suggest formation through core accretion.
However, its mass is significantly higher than the expected lower mass limit for deuterium burning \citep[$11-16$ $M_\text{J}$, ][]{mordasinietal2012}, which strongly points towards it being a deuterium-burning BD.

\subsection{Search for short-period companions}

\begin{figure}
    \centering
    \includegraphics[width=0.45\textwidth]{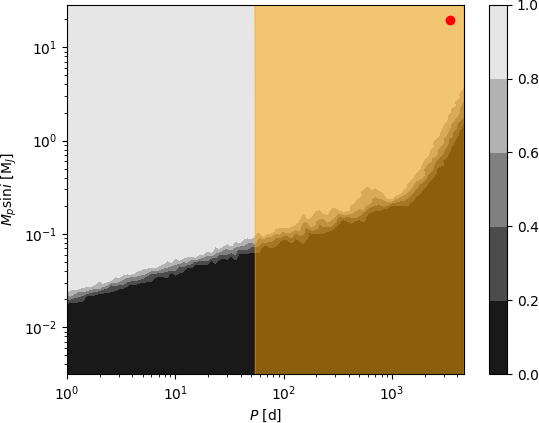}
    \caption{Detection map for the HARPS-N RV data of HD\,128717. The colour scale represents the detection function. The red point shows the position of HD\,128717\,B in the parameter space. The orange shaded area is the forbidden region due to the Hill stability criterion.}
    \label{fig:detection_map}
\end{figure}

In Sect. \ref{sec:add_rv}, we tried to fit additional signals in the RV time series to test the presence of additional objects in the system. We found no significant signal in the periodogram of the RV residuals and all other models showed a higher BIC than the single-Keplerian one; namely, no statistical evidence of a more complex model was found. Nonetheless, small planetary companions could still be undetected in the HARPS-N data. To quantify this, we computed the detection function of the RV data, following the same formalism adopted in \citet{ruggierietal2024} and \citet{sozzettietal2024}. The resulting detection map is shown in Fig. \ref{fig:detection_map}. It is worth noticing that low-mass super Earths ($M_p \sin{i} \lesssim 10$ $M_\oplus$) would not be detected in our time series, except at the shortest periods. 
This in spite of the high-precision, high cadence HARPS-N RV time-series, because of the significant levels of RV jitter.

However, due to the high mass and large eccentricity of Gaia-6\,B, many orbital configurations would be unstable. This means that we can rule out the presence of planetary companions even below the detectability threshold. We can compute the orbital stability region using Hill's Criterion \citep{hamiltonburns1992}. The region of the period-mass parameter space\footnote{Assuming edge-on orbit, $\sin i = 1$.} in which a planet on a circular orbit would be unstable under Hill's Criterion is highlighted in orange in Fig. \ref{fig:detection_map}. We can conclude that the only possible undetected planetary companions in the inner regions have $M_p \sin{i} \lesssim 10$ $M_\oplus$ and $P \lesssim 50$ d.

\subsection{Imaging follow-up observations}

    \begin{figure}
        \centering
        \includegraphics[width=\linewidth]{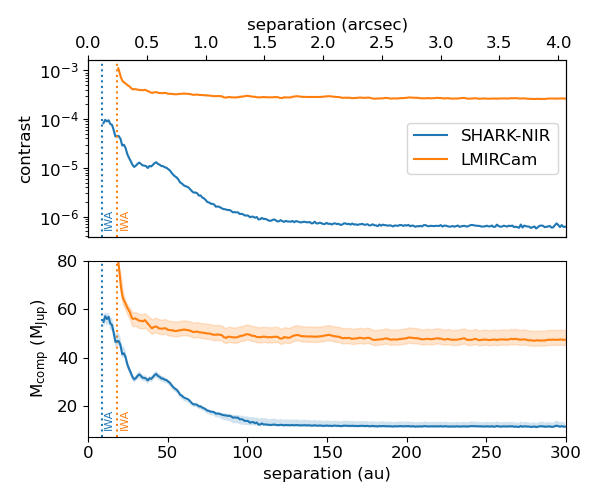}
        \caption{Contrast curve and mass limits for the SHARK-NIR (blue) and LMIR-Cam (orange) observations of HD\,128717 conducted on the night of May 16, 2025 UT. \textit{Top panel}: Contrast curve. \textit{Bottom panel}: Mass limits, with the thick curve corresponding to the nominal 1.40 Gyr stellar age and the shaded region to the age uncertainty.}
        \label{fig:contrast}
    \end{figure}
    The high eccentricity of HD\,128717\,B can be interpreted as the result of orbital excitation caused by a massive companion at large separation. As the orbital coverage provided by our HARPS-N RV data is too low to robustly detect any long-term trend caused by such a distant companion, we searched for any outer companion to HD\,128717 with imaging, a detection technique uniquely suited to discover massive companions on large orbits beyond the reach of RV observations.
    
    We observed HD\,128717 with SHARK-NIR \citep{farinato2022,marafatto2022} and LBTI/LMIRCam \citep{skrutskie2010,leisenring2012}, two infrared imagers mounted on the Large Binocular Telescope (LBT) in Arizona for multi-band simultaneous observations, on April 15 and May 16, 2025. Details of these observations are discussed in Appendix \ref{app:imaging}.
    In light of the longer duration and better seeing, in the following, we focus on the May 16 observation alone.
    
    No source with a S/N ratio higher than 4 is present on either of the post-processed images, allowing us to conclude that no companions were successfully detected around HD\,128717 during our observations. This non-detection can be used to provide robust limits on the mass of any yet-undetected outer companion in the system. Using the AMES-COND models \cite{allard2003} and the star host age of $1.4\pm0.3$ Gyr (computed as discussed in Appendix \ref{app:star_age}), we derived the mass detection limits for the imaging observations, shown in the bottom panel of Fig. \ref{fig:contrast}. While no low-mass GP companion would have been detectable in our data, our observations would have instead been able to robustly detect any large-separation stellar or BD companion in the system. Specifically, focussing on the mass limit computed for the nominal age of 1.4 Gyr and orbital separations beyond 60 au, LMIRCam observations in the L$^\prime$ band are able to exclude the presence of any outer companion more massive than $\sim 47$ $M_\text{Jup}$, while the SHARK-NIR data in the H band allow for the robust exclusion of any distant companion above $\sim 11$ $M_\text{Jup}$.

\section{Conclusion}
\label{sec:conclusion}

We have confirmed the sub-stellar nature of the astrometric candidate Gaia-ASOI-009, thanks to a high-cadence high-precision RV follow-up with the HARPS-N spectrograph. In brief, Gaia-6\,B is a high-eccentricity BD, $M_\text{B} = 19.8$ $M_\text{J}$, $e_\text{B} = 0.85$, and it is one of the most eccentric BDs with a precisely measured mass known so far (see Fig. \ref{fig:bd_ecc}). Moreover, we solved the apparent discrepancy between the orbital measurement from Gaia DR3 and the high-precision HARPS-N RV follow-up. This proves the importance of RV follow-up of astrometric candidates to confirm and improve the precision on the orbital parameters.

The origin of the high eccentricity of Gaia-6\,B remains an unsolved puzzle, since we did not identify any hint of the presence of other companions in the RV, astrometric, and direct imaging data analysed in this study. This is unexpected, since all known high-eccentricity ($>0.9$) planetary companions have clearly identified additional companions on wide orbits \citep[and references therein]{sozzettietal2023} and there is a known link between eccentricity and stellar multiplicity \citep[e.g.][]{moutouetal2017}.
However, the current data are not sufficient for us to conclusively exclude the presence of small BD companions at around a few tens of astronomical units, which might still be sufficient to excite Gaia-6\,B eccentricity to the observed level. In the future, additional observations, both additional RVs data to expand the temporal baseline and deeper imaging observations to search for smaller companions, will help tighten the constraints on the presence of undetected companions. Meanwhile, in-depth dynamical simulations might help shed light on the possible mechanisms producing such high eccentricity, which might be more common than expected among long-period GPs and BDs.

\section*{Data availability}
Full RV and activity HARPS-N data (Sect. \ref{sec:spec_data} and \ref{app:activity}) are only available in electronic form at the CDS via anonymous ftp to cdsarc.u-strasbg.fr (130.79.128.5) or via \url{http://cdsweb.u-strasbg.fr/cgi-bin/qcat?J/A+A/}.

\begin{acknowledgements}
We acknowledge support from the European Union – NextGenerationEU (PRIN MUR 2022 20229R43BH) and the ``Programma di Ricerca Fondamentale INAF 2023''. We acknowledge financial contribution from the INAF Large Grant 2023 ``EXODEMO''.
MPi acknowledges support from ASI-INAF agreement no. 2025-10-HH.0 ``Partecipazione Italiana al Gaia DPAC - Supporto alle attivit\`a di responsabilit\`a del team scientifico''.
TZi acknowledges support from CHEOPS ASI-INAF agreement n. 2019-29-HH.0, NVIDIA Academic Hardware Grant Program for the use of the Titan V GPU card and the Italian MUR Departments of Excellence grant 2023-2027 ``Quantum Frontiers''.
L.M. acknowledges financial contribution from PRIN MUR 2022 project 2022J4H55R.
We acknowledge the Italian center for Astronomical Archives (IA2, \href{https://www.ia2.inaf.it}{https://www.ia2.inaf.it}), part of the Italian National Institute for Astrophysics (INAF), for providing technical assistance, services and supporting activities of the GAPS collaboration.
We thank Tom Herbst of MPIA-Heidelberg (D) and the NIRVANA team for sharing with us part of the NIRVANA instrument control SW to control the motorised axis of SHARK-NIR, and NASA and the PI of JWST/NIRCam [26] Marcia Rieke for allowing us to use one of the NIRCam spare detectors as scientific detector for the SHARK-NIR scientific camera.
Eventually, we emphasise that observations have benefited from the use of ALTA Center (alta.arcetri.inaf.it) forecasts performed with the Astro-Meso-Nh model. Initialisation data of the ALTA automatic forecast system come from the General Circulation Model (HRES) of the European Centre for Medium Range Weather Forecasts.
This research has made use of the NASA Exoplanet Archive, which is operated by the California Institute of Technology, under contract with the National Aeronautics and Space Administration under the Exoplanet Exploration Program, and of data obtained from or tools provided by the portal exoplanet.eu of The Extrasolar Planets Encyclopaedia.
This work has made use of data from the European Space Agency (ESA) mission {\it Gaia} (\url{https://www.cosmos.esa.int/gaia}), processed by the {\it Gaia} Data Processing and Analysis Consortium (DPAC, \url{https://www.cosmos.esa.int/web/gaia/dpac/consortium}). Funding for the DPAC has been provided by national institutions, in particular the institutions participating in the {\it Gaia} Multilateral Agreement.

\end{acknowledgements}

\bibliographystyle{aa} 
\bibliography{biblio} 

\begin{appendix}

\section{Additional activity analysis}
\label{app:activity}

\begin{figure}
   \centering
   \includegraphics[width=0.45\textwidth]{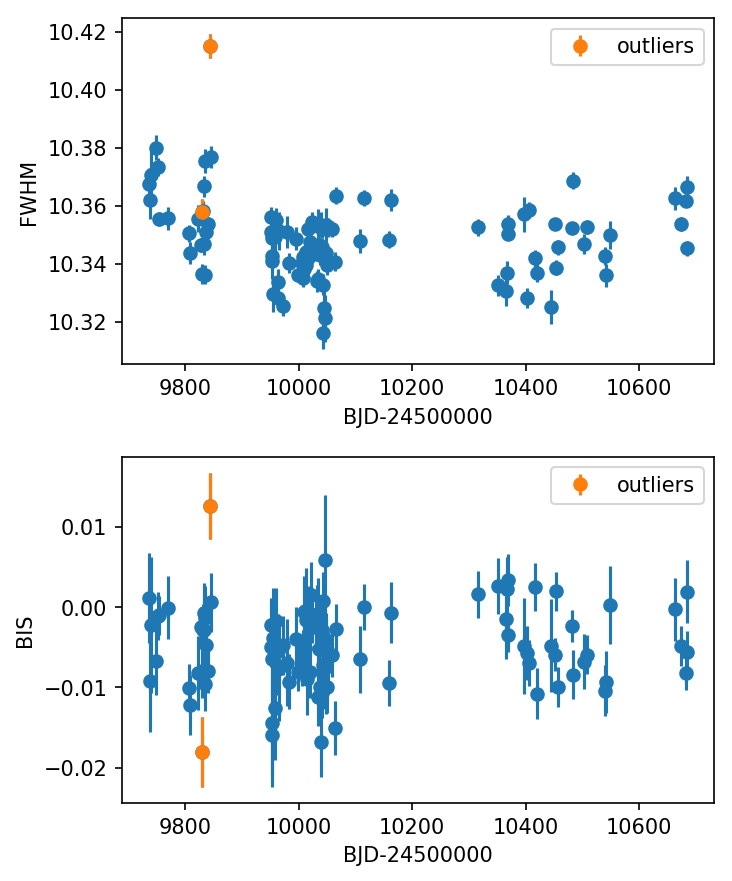}
 \caption{Time series of the HARPS-N CCF activity indicators. \textit{Upper panel:} FWHM; \textit{Lower panel:} BIS. The orange dots represents the outliers discussed in Sect. \ref{sec:1_plan}.}
 \label{fig:ccf_timeseries}
\end{figure}

In addition to the $\log R'_\text{HK}$ analysis, we studied two other activity indicators, derived from the CCF of the HARPS-N spectra via version 3.0.1 of the HARPS-N Data Reduction Software \citep[DRS,][]{dumusqueetal2021}: the full width half maximum (FWHM) and bisector span (BIS). The two time series are shown in Fig. \ref{fig:ccf_timeseries}. Of the two outliers discussed in Sect. \ref{sec:1_plan}, one can be clear seen as an outlier in both CCF activity time series.

\begin{figure}
   \centering
 \includegraphics[width=.45\textwidth]{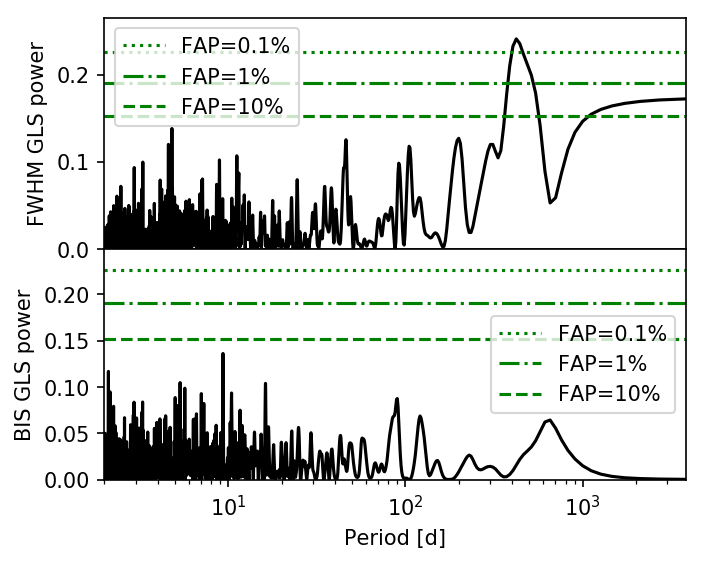}
 \caption{GLS periodograms of the HARPS-N CCF activity indicators. \textit{Upper panel:} FWHM; \textit{Lower panel:} BIS.}
 \label{fig:periodograms_ccf}
\end{figure}

In Fig. \ref{fig:periodograms_ccf}, we show the GLS periodograms of the two time series. The FWHM shows a significant peak at $P= 422 \pm 27$ d, close to the one highlighted in the $\log R'_\text{HK}$ time series (see Fig. \ref{fig:periodograms_rv_rhk}), while the BIS does not show any significant periodicity above the FAP$=10\%$ level.

\section{Gaia DR3 priors model}
\label{app:gaia_emcee}
As an additional test to attempt reconciling the results from Gaia astrometry and HARPS-N RVs, we performed an MCMC fit of the RV time series using Gaussian priors from the Gaia orbital solutions, using the best-fit values and their uncertainties. The adopted values of these priors are listed in Table \ref{tab:mcmc_gaia_param}. It is worth noticing that while some parameters come directly from the orbital solutions published in the Gaia archive, such as those on $P_\text{b}$ and $T_{p,b}$, the others were derived from the related parameters and corresponding uncertainties (see Sect. \ref{sec:spec_data}). One particular case is that of the semi-amplitude $K_\text{b}$, which directly depends on the companion mass, for which no uncertainty was provided in the Gaia DR3 catalogue: for this reason, we did not fix a Gaussian prior on $K_\text{b}$ and adopted the uninformative prior from Table \ref{tab:mcmc_prior_param}. The results of this fit are listed in the last columns of Table \ref{tab:mcmc_gaia_param}: there is no significant deviation from the results of the original 1-Keplerian fit with uninformative priors from Table \ref{tab:mcmc_prior_param}. This suggests that the evidence from the RV time series is stronger than the constraint from the astrometric orbital solution, as expected.

\begin{table}
\caption[]{Priors and best-fit parameters for the tested Gaia-constrained MCMC model.}
\label{tab:mcmc_gaia_param}
\centering
\begin{tabular}{lcc}
\hline
\hline
\noalign{\smallskip}
& Priors   & Values   \\
\hline
\noalign{\smallskip} 
$K_\text{b}$ $[$m s$^{-1}]$ & $\mathcal{U}$(0,1000) & $322.2^{+3.7}_{-3.5}$ \\ 
\noalign{\smallskip}  
$P_\text{b}$ $[$d$]$ & $\mathcal{N}$(1090, 310) & $2513^{+72}_{-69}$ \\ 
\noalign{\smallskip}  
$T_{p,b}$ $[$BJD$-2450000]$ & $\mathcal{N}$(10490, 290) & $10159.07^{+0.58}_{-0.60}$ \\
\noalign{\smallskip}  
$\sqrt{e_\text{b}} \cos{\omega_\text{b}}$ & $\mathcal{N}$(0.28, 0.32) & $-0.229^{+0.011}_{-0.011}$ \\
\noalign{\smallskip}
$\sqrt{e_\text{b}} \sin{\omega_\text{b}}$ & $\mathcal{N}$(-0.22, 0.41) & $-0.8646^{+0.0048}_{-0.0048}$ \\
\noalign{\smallskip}
\hline
\noalign{\smallskip}  
$M_\text{b} \sin i$ $[M_\text{J}]$  &  & $14.70^{+0.50}_{-0.51}$ \\ 
\noalign{\smallskip}  
$a_\text{b}$ $[$AU$]$  & & $3.860^{+0.099}_{-0.097}$ \\ 
\noalign{\smallskip}  
$e_\text{b}$  & & $0.8002^{+0.0063}_{-0.0065}$ \\ 
\noalign{\smallskip}  
$\omega_{\star,b}$ $[$rad$]$ & & $-1.830^{+0.013}_{-0.013}$ \\ 
\noalign{\smallskip}  
\hline  
\noalign{\smallskip}  
$\gamma_\text{HARPS-N}$ $[$m s$^{-1}]$ & $\mathcal{U}$(-50.0,250.0) & $237.49^{+0.96}_{-0.95}$ \\ 
\noalign{\smallskip}  
$\sigma_\text{jit,HARPS-N}$ $[$m s$^{-1}]$ & $\mathcal{U}$(0,100) & $5.25{+0.44}_{-0.39}$ \\ 
\noalign{\smallskip}
\hline
\end{tabular}
\end{table}

\section{Stellar age and rotation period}
\label{app:star_age}

As discussed in Sect. \ref{sec:star_par}, the isochrones estimate of the stellar age points towards an intermediate age, although with large uncertainties, $t_\text{iso} = 1.9^{+1.9}_{-1.3}$ Gyr. We consider here additional age diagnostics to further constrain the system age.

In Sect. \ref{sec:star_par}, we highlight the presence of a significant Lithium line in the spectra of HD\,128717. The abundance of Lithium is commonly associated with a young stellar ag, as Lithium depletes over time in main-sequence stars. However, comparing HD\,128717 Lithium equivalent width and abundance, and its effective temperature with those of clusters of known age, the parameters of HD\,128717 are in good agreement with the distribution of the open cluster NGC 752, which has a measured age of 1.34 Gyr \citep{aguerosetal2018}.

Another common proxy for stellar age is the activity level: the mean $\log R'_\text{HK} = -4.947$ is quite low. Following activity-rotation and gyrochronology relations \citep{noyesetal1984,mamajekhillenbrand2008}, we can derive a rotation period of $P_\text{rot} = 21 \pm 4$ d and an age of $t = 3.7 \pm 0.4$ Gyr. However, this rotation period is incompatible with the measured projected rotational velocity, $v \sin i = 6.1 \pm 0.5$ km s$^{-1}$ (in good agreement with the $v \sin i$ $T_\text{eff}$ relation by \citet{winnetal2017}), that combined with the stellar radius $R_\star=1.248 \pm 0.024$ $R_\odot$ results in a maximum rotation period of $P_\text{rot, max} = 10.3 \pm 0.9$ d. 

To obtain a more precise estimate of the rotation period, we analysed 4 available TESS sectors (14,15,49 and 50) with the Python wrapper \texttt{juliet}\footnote{\url{https://juliet.readthedocs.io/en/latest/}} \citep{espinozaetal2019} using a quasi-periodic Gaussian process (GP) kernel implemented with \texttt{celerite} \citep{foremanmackeyetal2017}. The GP modelling converges on a period of $3.9 \pm 0.3$ d. 

If we interpreted this as the stellar rotation period, it would point towards a young age ($\simeq 300$ Myr), in contrast with the other indicators. However, the peak in the GP posterior could correspond to the first harmonic of the stellar rotation, resulting in $P_\text{rot} = 7.8 \pm 0.6$ d, which would correspond in a more compatible estimate of gyrochronological age, $t_\text{gyro} = 900$ Myr. 

Finally, we were able to obtain another estimate of the age of the system from its kinematic. We can compute the Galactic space velocity components $U$, $V$, and $W$ from the proper motion, parallax, and radial velocity of the system. We can get the Gaia DR3 parallax from Table \ref{tab:star_par} and, from the HARPS-N spectra collected, we can derive the absolute systemic velocity of the star using version 3.0.1 of the HARPS-N Data Reduction Software \citep[DRS,][]{dumusqueetal2021}: $\gamma_\text{syst} = + 5.90776 \pm 0.00101$ km s$^{-1}$. We do not use the Gaia DR3 proper motion components, as the measurements can be biased by the incorrect modelling of the Keplerian motion. For this reason, we adopt the proper motion from \citet{Brandt2021}, which is computed from the difference in position between the Gaia DR3 and Hipparcos catalogues, and therefore should not be affected by the orbital motion\footnote{The Gaia DR3 - Hipparcos proper motion could still be affected by additional undetected companions: although massive long-period companions undetected by the direct imaging observations (Fig. \ref{fig:contrast}) could still produce significant astrometric distortion, this would not significantly affect the estimate of the kinematical age of the system.}: $\mu_\text{ra} = -81.992 \pm 0.021$ mas yr$^{-1}$, $\mu_\text{dec} = +67.612 \pm 0.023$ mas yr$^{-1}$. From these values, following the methodology by \citet{johnsonsoderblom1987}, we obtain: $U = -35.876 \pm 0.059$ km s$^{-1}$, $V= -10.698 \pm 0.013$ km s$^{-1}$, $W = -3.673 \pm 0.005$ km s$^{-1}$. From these Galactic space velocities, we can compute the kinematical age of the system, as defined by \citet{almeidafernandes2018}, as $t_\text{kin} \simeq 1.4$ Gyr. It is worth mentioning that the kinematical age gives only an approximate estimate for singles systems, and it is more informative for statistical samples, and in fact \citet{almeidafernandes2018} suggest an uncertainty of around 3 Gyr for individual stars. However, it is noteworthy that this value is in good agreement with the previous estimates. Combining all these estimates, we can adopt the average value of the age for HD\,128717 $t = 1.4 \pm 0.3$ Gyr.

\section{Direct imaging observations and analysis}
\label{app:imaging}
In both observation epochs SHARK-NIR used its board-band H filter and Gaussian coronagraph with an inner working angle (IWA) of 120 mas, while LMIRCam observations were performed in the L$^\prime$ band using the vector-apodising phase plate coronagraph \citep[vAPP, ][]{doelman2021} having an IWA of 246 mas. The coronagraphs IWAs correspond for HD\,128717 to orbital separations of $\sim$8 and $\sim$18 au, meaning that HD\,128717\,B would remain covered by the central occulter even at apoastron and therefore undetectable. The April observations lasted $\sim$40 minutes and where characterised by rapidly variable seeing between 1.09$^{\prime\prime}$ and 2.30$^{\prime\prime}$, while the May epoch lasted $\sim$1 hour with seeing between 0.90$^{\prime\prime}$ and 1.49$^{\prime\prime}$.

The SHARK-NIR and LMIRCam data thus collected were reduced using a post-processing procedure based on the angular differential imaging method \citep[ADI, ][]{marois2006} method and the principal components analysis \citep[PCA][]{soummer2012} algorithm. For both instruments, contrast limits were computed by estimating the standard deviation within one-pixel-wide annuli centred on the star, with the resulting contrast curves shown in the upper panel of Fig. \ref{fig:contrast}.

\section{BD and GP statistical sample}
\label{app:bd_gp_pop}

\begin{figure}
   \centering
   \includegraphics[width=0.45\textwidth]{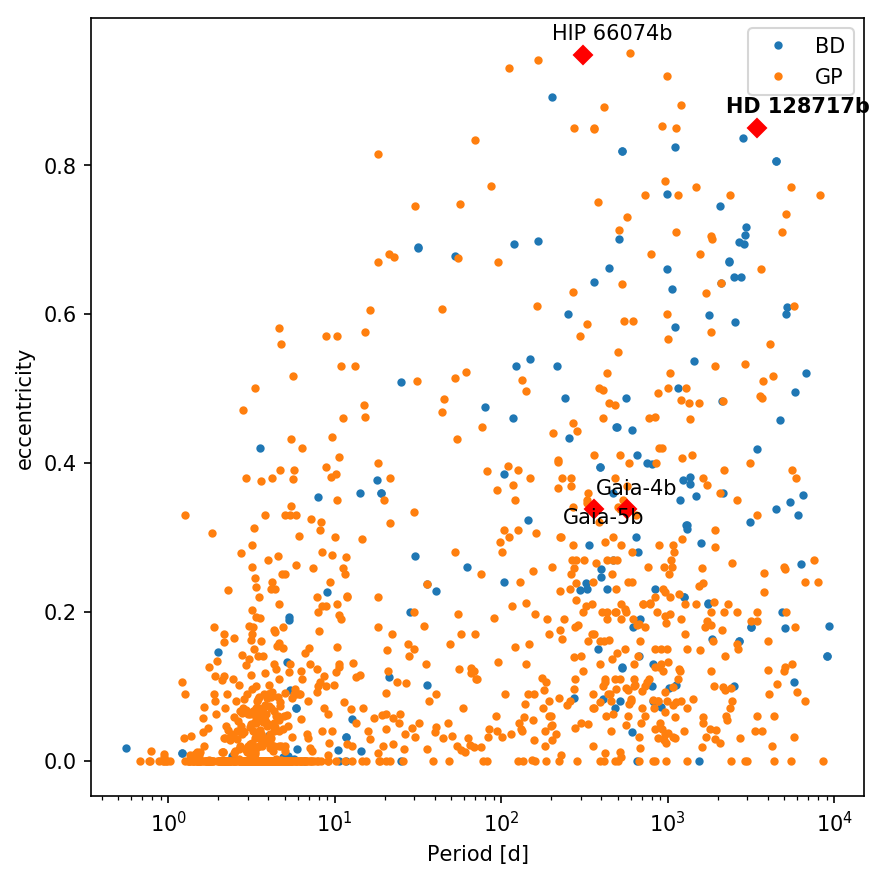}
 \caption{Eccentricity of GPs (orange) and BDs (blue) as a function of the orbital period, for substellar companions with period $<10^4$ d. The red diamonds highlight the positions of the confirmed Gaia-ASOI sub-stellar companions and HIP 66074.}
 \label{fig:bd_ecc}
\end{figure}

To compare Gaia-6\,B with the known population of BDs and GPs, we collected from the public archives a large sample of well characterised systems with BD and GP companions. First we selected from the NASA Exoplanet Archive\footnote{\url{https://exoplanetarchive.ipac.caltech.edu/}} all objects with mass (or $m \sin i$) $> 0.1$ $M_\text{J}$ and period $< 10^4$ d, considering only the default solutions from the archive. Then we selected only the systems with a measured stellar mass and metallicity, resulting in 1138 companions. We then selected from the Encyclopaedia of Exoplanetary Systems\footnote{\url{https://exoplanet.eu/}} all BDs with mass (or $m \sin i$) $> 13.$ $M_\text{J}$, selecting again only companions with period $< 10^4$ d and measured mass and metallicity of the stellar host, corresponding to 78 companions. We then merged this sample with the catalogue of BD companions from Gaia DR3 selected by \citet{stevensonetal2023} which, after applying the same selection as for the other two catalogues, contains 117 objects.
Merging this three samples together, discarding duplicates, we obtain a sample of 1311 companions, with 1118 and 193 GPs and BDs, respectively. Fig. \ref{fig:brown_dwarfs} shows the distribution of companion mass as a function of the stellar mass and metallicity, while Fig. \ref{fig:bd_ecc} shows the period-eccentricity distribution of GPs and BDs. In both figures are highlighted Gaia-4,-5, and -6\,B as well as HIP66074\,b.

\section{\texttt{emcee} auxiliary plots}
\label{app:emcee_plots}

\begin{figure*}
   \centering
   \includegraphics[width=0.9\textwidth]{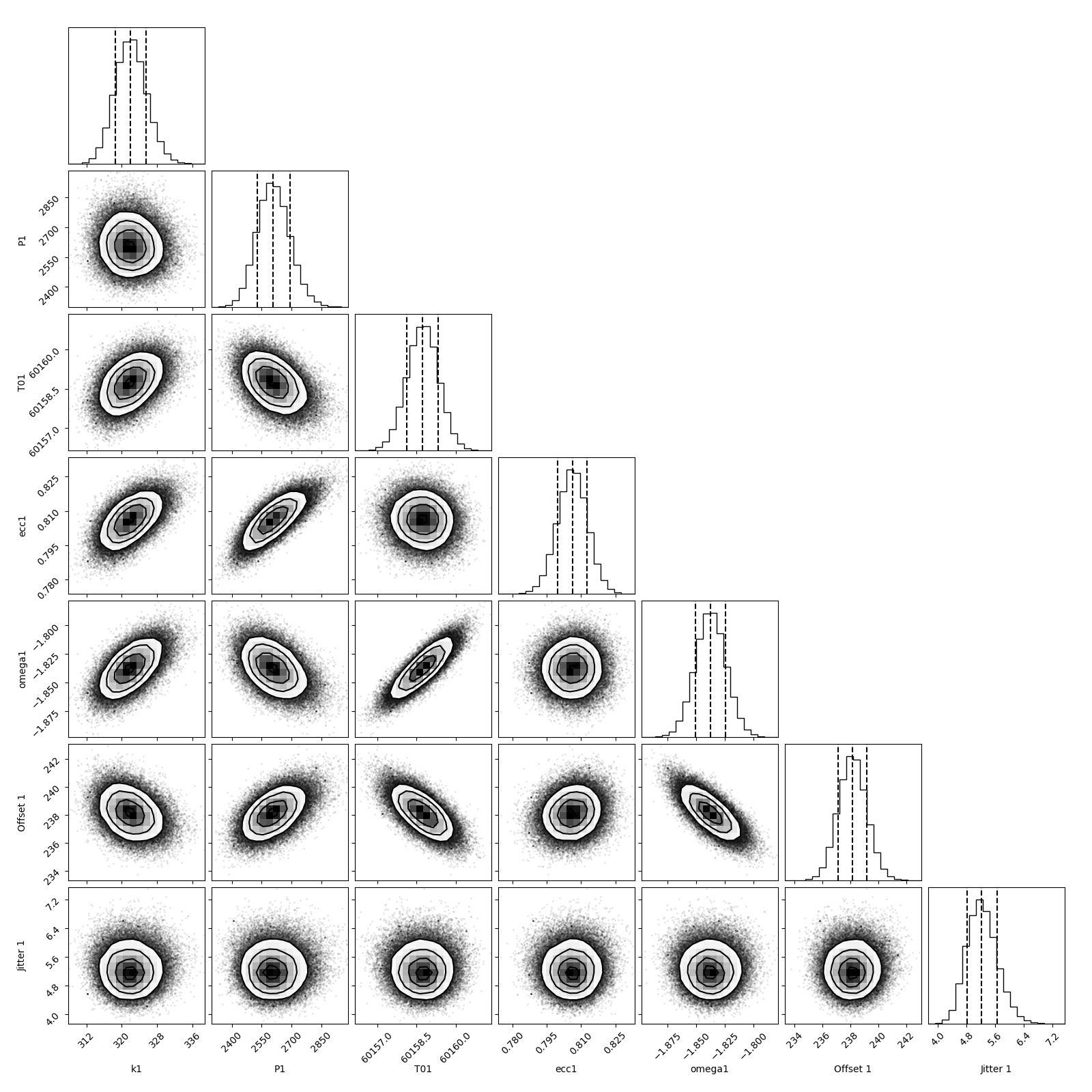}
 \caption{Posterior distributions of the fitted parameters of the 1-Keplerian model of the HARPS-N RV time series. The vertical dashed lines denote the median and the 16th and 84th percentiles. The shown parameters are, in order: $k_\text{b}$, $P_\text{b}$, $T_\text{p,b}$, $e_\text{b}$, $\omega_\text{b}$, $\gamma_\text{HARPS-N}$, and $\sigma_\text{jit,HARPS-N}$}
 \label{fig:one_plan_posterior}
\end{figure*}

Fig. \ref{fig:one_plan_posterior} shows the complete posteriors of the 1-Keplerian model of the HARPS-N RV data, which was adopted as the best RV model.

\end{appendix}

\end{document}